\documentclass[14pt]{ap-jnmp}
\usepackage{color}

\usepackage{amsmath,amssymb}
\usepackage{graphicx}
\usepackage[justification=centering,font=small]{caption}

\usepackage{titlesec}
\titleformat{\subsection}
{\normalfont\fontsize{12}{17}\bfseries\itshape}{\thesubsection}{1em}{}

\usepackage{hyperref}

\def\dtp#1{\mathop {#1}\limits_{+\tau}}
\def\dtm#1{\mathop {#1}\limits_{-\tau}}

\def\dsm#1{\mathop {#1}\limits_{-s}}

\def\sech{\,\textrm{sech}}

\copyrightauthor{}

\def\bib{B\kern-.05em{I}\kern-.025em{B}\kern-.08em}
\def\btex{B\kern-.05em{I}\kern-.025em{B}\kern-.08em\TeX}

\title{Symmetries, conservation laws, invariant solutions and
   difference schemes
  \\
   of the one-dimensional Green-Naghdi equations}

\author{V.A. DORODNITSYN$^a$, E.I. KAPTSOV$^{a,b}$ and  S.V. MELESHKO$^b$, \footnote{Corresponding author.}}

\address{$^a$ Keldysh Institute of Applied Mathematics, Russian Academy of Science, \\
Miusskaya Pl.~4, Moscow, 125047, Russia;
\\
\email{Dorodnitsyn@Keldysh.ru}}

\address{$^b$ School of Mathematics, Institute of Science, \\
 Suranaree University of Technology, 30000, Thailand\\
\email{evgkaptsov@gmail.com}
\email{sergey@math.sut.ac.th}}

\begin{document}

\maketitle
\thispagestyle{empty}

\vphantom{\vbox{%
\begin{history}
\received{(Day Month Year)}
\revised{(Day Month Year)}
\accepted{(Day Month Year)}
\end{history}
}}

ABSTRACT

\begin{abstract}
The paper is devoted to  the Lie group properties of the
one-dimensional Green-Naghdi equations describing the  behavior
of fluid flow over uneven bottom topography. The bottom topography
is incorporated into the Green-Naghdi equations in two ways: in the
classical Green-Naghdi form and in the approximated form of the same
order. The study is performed in Lagrangian coordinates which
allows one to find Lagrangians for the analyzed equations. Complete
group classification of both cases of the Green-Naghdi equations
with respect to the bottom  topography is presented.
Applying Noether's theorem, the  obtained
Lagrangians and the group classification, conservation laws of the
one-dimensional Green-Naghdi equations with uneven bottom topography
are obtained. Difference schemes which preserve the symmetries of
the original equations and the conservation laws are constructed.
Analysis of the developed schemes is given. The schemes are tested
numerically on the example of an exact traveling-wave
solution.

\end{abstract}



\section{Introduction}

An ideal fluid flows under the force of gravity can be modeled
 by means of the Euler equations. However, the full Euler equations
are too complicated for describing waves on the surfaces of ideal
fluid, in particular, because of free surface being a
part of the solution. This difficulty motivated scientists for
deriving simpler equations. One class of such equations is a class
of shallow water equations. The classical approach of deriving the shallow
water equations consists of approximation of the Euler equations for
the irrotational flows. The hierarchy of the shallow water
approximations is considered with respect to the shallowness
parameter $\delta=h_{0}/L$, where $h_{0}$ is the mean depth of the
fluid, $L$ is the typical length scale of the wave
\cite{bk:Matsuno[2015]}\footnote{See also the references therein.}.
In particular, the Green-Naghdi equations, derived for
describing the two-dimensional  fluid flow over an uneven bottom,
are accurate to the dispersive terms of order $\delta^{2}$. The
Green-Naghdi system of equations is the generalization of the
equations derived first by Serre \cite{bk:Serre[1953]} and later by
Su and Garden \cite{bk:SuGardner[1969]} to describe the
one-dimensional propagation of fully nonlinear and weakly dispersive
surface gravity waves over flat bottom. In the present paper we
study the one-dimensional case of the Green-Naghdi equations.

Due to the significance of the Green-Naghdi model, there has been
increasing interest in the numerical solving of the Green-Naghdi equations.
The following numerical approaches have been developed for the Green-Naghdi
system, based on either the finite difference, hybrid finite-difference
finite-volume, pseudospectral and Galerkin/finite-element methods.
Review of these methods can be found in \cite{bk:Bonneton_etal2011,bk:DutykhClamondMilewskiMitsotakis2013,%
bk:MitsotakisIlanDutykh2014,bk:LiXuCheng2019}.
A numerical scheme developed in \cite{bk:FavrieGavrilyuk2017} is based
on a different approach: the idea is to replace the dispersive Green-Naghdi
equations by approximate hyperbolic equations. This approach was also applied
in \cite{bk:LiapidevskiiGavrilova2008,bk:GavrilyukLiapidevskiiChesnokov2016}.

In the present paper we use an approach based on the group
properties of the Green-Naghdi equations. The group analysis method
\cite{bk:Ovsiannikov1978,bk:Olver[1986]} yields  exact solutions of
differential equations, conservation laws via Noether's theorem and
a basis for invariant finite-difference schemes construction.
Applications of group analysis to the Green-Naghdi equations with a horizontal
bottom topography in
Eulerian and Lagrangian coordinates were studied in
\cite{bk:BagderinaChupakhin[2005],bk:SiriwatKaewmaneeMeleshko2016}.
In particular, the authors of \cite{bk:SiriwatKaewmaneeMeleshko2016}
applied Nother's theorem for finding conservation laws of the
Green-Naghdi equations. Notice that in order to apply the Noether
theorem for finding conservation laws one needs to know an admitted
Lie group. Another requirement is an existence of a Lagrangian
providing that the underline equations are in the Euler-Lagrange
form. This is one of the main advantages of the Lagrangian coordinates
used in the present paper.

One of the main objectives of the present paper among the group classification
and conservation laws is to
construct a numerical scheme which inherits the group properties of
the original equations.
Earlier, in \cite{bk:DorKapSW2019,bk:DorKapSW2020}, this method was applied to the hyperbolic shallow water equations  with an arbitrary bottom topography considered in Eulerian coordinates.
An invariant difference scheme possessing all difference analogues
of the conservation laws was constructed there.


The one-dimensional Green-Naghdi equations describing
surface gravity waves over uneven bottom have the form \cite{bk:GreenNaghdi[1976],Bazdenkov_C_1987,bk:LannesBonneton2009,bk:Matsuno[2016]}
\begin{equation}
\begin{array}{c}
{\displaystyle \frac{\partial}{\partial t}h+\frac{\partial}{\partial x}(hu)=0,}
\\[2ex]
{\displaystyle \frac{d}{dt}u+g\frac{\partial}{\partial x}(h+H_{b})=\frac{\epsilon}{h}\left(\frac{\partial}{\partial x}A+B\frac{\partial}{\partial x}H_{b}\right),}
\end{array}\label{eq:SGN}
\end{equation}
where $h$ is the depth of the layer of fluid, $u$ is the velocity,
$t$ is time, $x$ is the Eulerian coordinate, $H_{b}(x)$ is the
function describing the bottom topography, $\epsilon=\delta^{2}$,
${\displaystyle \frac{d}{dt}=\frac{\partial}{\partial t}+u\frac{\partial}{\partial x}}$
is the material derivative, and
\begin{equation}
{\displaystyle A=h^{2}\frac{d}{dt}\left(\frac{h}{3}u_{x}-\frac{1}{2}uH_{bx}\right),\,\,\,B=h\frac{d}{dt}\left(\frac{h}{2}u_{x}-uH_{bx}\right),}\label{eq:original_bottom}
\end{equation}

Assuming that
\[
H_{b}(x)=H_{b0}+\epsilon D(x),\,\,\,(H_{b0}=\textrm{const}),
\]
or
\[
H_{b}(x)=D(\epsilon x),
\]
and cutting terms of order $O(\epsilon^{2})$, one obtains equations (\ref{eq:SGN}) with the mild slope approximation, where
\begin{equation}
A=h^{2}\frac{d}{dt}\left(\frac{h}{3}u_{x}\right),\,\,\,B=0.\label{eq:simplified_bottom}
\end{equation}

It should be noted here that for $\varepsilon=0$ equations (\ref{eq:SGN}) become
the hyperbolic shallow water equations. Symmetries, conservation laws and numerical
schemes based on their group properties of the one-dimensional hyperbolic shallow
water equations with different type of bottom topographies have been  analyzed in
Eulerian and Lagrangian coordinates in  \cite{bk:BagderinaChupakhin[2005],bk:SiriwatKaewmaneeMeleshko2016,%
bk:KaptsovMeleshko2019_2,%
bk:AksenovDruzhkov2016b,bk:AksenovDruzhkov[2019],bk:AksenovDruzhkov[2020],%
bk:DorKapSW2019,bk:DorodnitsynKozlovMeleshko2019,bk:DorKapSW2020}.

In the present paper the one-dimensional equations (\ref{eq:SGN})
in mass Lagrangian coordinates $(s,t)$ with the functions $A$ and
$B$ of the form either (\ref{eq:original_bottom}) or (\ref{eq:simplified_bottom})
are studied.
In \cite{bk:SiriwatKaewmaneeMeleshko2016}, it is shown that the
Green-Naghdi equations with horizontal plane bottom are equivalent
to the Euler-Lagrange equation written in the mass Lagrangian coordinates.
For the one-dimensional
Green-Naghdi equations with uneven bottom there is the problem of finding a Lagrangian. If the Lagrangian
is found, then one can apply Noether's theorem for finding conservation
laws.

Another objective of the present paper is to construct conservative
numerical schemes which preserve both a symmetry of original differential
equations and difference analogs of conservation laws.

\medskip

The paper is organized as follows.  Section 2 is devoted to the
study of equations (\ref{eq:SGN}), (\ref{eq:original_bottom}) in
Lagrangian coordinates:  corresponding Lagrangians are found, and
the group classification is carried out. Applying Noether's theorem,
conservation laws in Lagrangian and Eulerian coordinates are
derived. Then, the similar study of equations (\ref{eq:SGN}),
(\ref{eq:simplified_bottom}) is given in Section 3, where it is also
shown that equations (\ref{eq:SGN}), (\ref{eq:simplified_bottom})
with a flat bottom topography $H_{b}=qx+\beta$ are locally
equivalent to the Green-Naghdi equations with a horizontal bottom
topography $H_{b}=\textrm{const}$. Preliminary information essential
for further construction of conservative finite-difference schemes
is given in Section~4. A new three-layer invariant conservative
finite-difference scheme for the Green-Naghdi equations with a
horizontal flat bottom topography is constructed in Section~5. At
the end of Section 5, the possibilities of extending the obtained
difference scheme to the case of an arbitrary bottom profile are
discussed. Application of the scheme for analysis of traveling
wave type solutions of the Green-Naghdi equations are considered in
Section~6. The results are summarized in Conclusion.

\section{Equations (\ref{eq:SGN}), (\ref{eq:original_bottom}) in Lagrangian
coordinates}

\subsection{Eulerian and Lagrangian coordinates}

Relations between Lagrangian coordinates $(t,\xi)$ and Eulerian coordinates
$(t,x)$ for the one-dimensional case are defined by the condition
$x=\varphi(t,\xi)$, where the function $\varphi(t,\xi)$ is the solution
of the Cauchy problem
\begin{equation}\label{lagr_mass_rel}
\varphi_{t}(t,\xi)=u(t,\varphi(t,\xi)),\,\,\,\varphi(t_{0},\xi)=\xi.
\end{equation}
In Lagrangian coordinates, the general solution of the mass conservation
law equation is
\[
h(t,\varphi(t,\xi))=\frac{h_{0}(\xi)}{\varphi_{\xi}(t,\xi)},
\]
where $h_{0}(\xi)$ is an arbitrary function of the integration such
that $h(t_{0},\xi)=h_{0}(\xi)$. Introducing the mass Lagrangian coordinate
$s$ \cite{bk:RozhdYanenko[1978]} by the equation
\begin{equation}
\xi=\alpha(s),\label{eq:change}
\end{equation}
where $\alpha^{\,\prime}(s)=h_{0}(\alpha(s))$, one obtains that in
the mass Lagrangian coordinates $(t,s)$
\[
\tilde{h}(t,s)\equiv h(t,\varphi(t,\alpha(s)))=\frac{1}{\tilde{\varphi}_{s}(t,s)}.
\]
Here the functions $\tilde{\varphi}(t,s)$ and $\varphi(t,\xi)$ are
related by the formula
\[
\tilde{\varphi}(t,s)=\varphi(t,\alpha(s)).
\]
Hence, the mass Lagrangian coordinates are defined by the equations
\[
\tilde{\varphi}_{t}(s,t)=\tilde{u}(s,t),\ \ \tilde{\varphi}_{s}(s,t)=\tilde{h}^{-1}(s,t).
\]
where $\tilde{u}(s,t)=u(\varphi(\alpha(s),t),t)$. The sign tilde
$\tilde{}\ $ is further omitted.  In order to derive representations
of the Green-Naghdi equations in the mass Lagrangian coordinates one
can use the following relations:
\[
\begin{array}{c}
u_{x}=\varphi_{ts}h,\,\,\,u_{t}=\varphi_{tt}-uu_{x},\,\,\,h_{x}=-\varphi_{ss}h^{3},\\[1.5ex]
u_{xx}=\varphi_{tss}+u_{x}h_{x}h^{-1},\,\,\,u_{tx}=\varphi_{tts}h-(uu_{xx}+u_{x}^{2}),\\[1.5ex]
u_{tt}=\varphi_{ttt}-(u^{2}u_{xx}+2uu_{tx}+uu_{x}^{2}+u_{x}u_{t}),\,\,\,h_{xx}=-\varphi_{sss}h^{4}+3h_{x}^{2}h^{-1},\\[1.5ex]
u_{txx}=\varphi_{ttss}h^{2}-\left(uu_{xxx}+3u_{x}u_{xx}-h^{-1}(uu_{xx}+u_{x}^{2}+u_{tx})h_{x}\right).
\end{array}
\]
The corresponding equation becomes
\begin{equation}
\begin{array}{c}
\epsilon(-3H_{b}^{\prime\prime\prime}\varphi_{t}^{2}\varphi_{s}^{5}-6H_{b}^{\prime\prime}H_{b}^{\prime}\varphi_{t}^{2}\varphi_{s}^{6}+6H_{b}^{\prime\prime}\varphi_{t}^{2}\varphi_{s}^{3}\varphi_{ss}-6H_{b}^{\prime\prime}\varphi_{t}\varphi_{ts}\varphi_{s}^{4}-3H_{b}^{\prime\prime}\varphi_{tt}\varphi_{s}^{5}-6H_{b}^{\prime}{}^{2}\varphi_{tt}\varphi_{s}^{6}\\
+6H_{b}^{\prime}\varphi_{tt}\varphi_{s}^{3}\varphi_{ss}-6H_{b}^{\prime}\varphi_{ts}^{2}\varphi_{s}^{3}-8\varphi_{tts}\varphi_{s}\varphi_{ss}+2\varphi_{ttss}\varphi_{s}^{2}+20\varphi_{ts}^{2}\varphi_{ss}-8\varphi_{ts}\varphi_{tss}\varphi_{s})\\
+6\varphi_{s}^{3}(-H_{b}^{\prime}\varphi_{s}^{3}g-\varphi_{tt}\varphi_{s}^{3}+\varphi_{ss}g)=0.
\end{array}\label{eq:june19.1}
\end{equation}

\subsection{Search for a Lagrangian}

For finding a Lagrangian for which equation (\ref{eq:june19.1}) is
the Euler-Lagrange equation one has to solve the following problem.
Let ${\cal
L}(t,s,\varphi_{t},\varphi_{s},\varphi_{tt},\varphi_{ts},\varphi_{ss})$
be a corresponding Lagrangian. Then,
 substituting ${\cal L}$ into
the equation\footnote{Notations related with the variational analysis follow to the formulations given in \cite{bk:Ibragimov2011}.}
 ${\displaystyle \frac{\delta{\cal
L}}{\delta\varphi}=0}$, excluding the derivative $\varphi_{ttss}$
found from equation (\ref{eq:june19.1}), and splitting it with
respect to the parametric derivatives
\[
\varphi_{ttt},\,\,\,\varphi_{tts},\,\,\,\varphi_{tss},\,\,\,\varphi_{sss},\,\,\,\varphi_{tttt},\,\,\,\varphi_{ttts},\,\,\,\varphi_{tsss},\,\,\,\varphi_{ssss},
\]
one obtains an overdetermined system of equations for the function
${\cal L}$. The general solution of these equations such that $2{\cal L}_{\varphi_{tt}\varphi_{ss}}+{\cal L}_{\varphi_{ts}\varphi_{ts}}\neq0$
gives the sought Lagrangian. Calculations, performed in symbolic manipulation system Reduce \cite{bk:Hearn}, give that the general solution
contains several arbitrary constants and unknown functions. These
functions satisfy a compatible system of partial differential equations.
As the general form of the Lagrangian is cumbersome, we only present here
a particular case of the Lagrangian
\[
{\cal L}=\frac{1}{2}\varphi_{t}^{2}\left(1+\epsilon(H_{b}^{\prime}{}^{2}-H_{b}^{\prime}\varphi_{s}^{-3}\varphi_{ss}+\frac{1}{2}H_{b}^{\prime\prime}\varphi_{s}^{-1})\right)+\frac{1}{6}\varphi_{s}^{-4}\left(\epsilon\varphi_{ts}^{2}-3g\varphi_{s}^{2}(2Q_{b}\varphi_{ss}+\varphi_{s})\right).
\]
where $Q_{b}^{\prime}(\varphi)=H_{b}(\varphi)$.

\subsection{Group analysis of equation (\ref{eq:june19.1})}

To find equivalence transformations we used the infinitesimal
criterion \cite{bk:Ovsiannikov1978}. For this purpose the
determining equations for the components of generators of
one-parameter groups of equivalence transformations were derived.
The solution of these determining equations gives the general form
of elements of the equivalence algebra of the class
(\ref{eq:june19.1}). The basis elements of the equivalence algebra
of the class (\ref{eq:june19.1}) are
\[
X_{1}^{e}=\varphi\partial_{\varphi}+4s\partial_{s}+t\partial_{t}+2H_{b}\partial_{H_{b}},\,\,\,X_{2}^{e}=\partial_{H_{b}},
\]
\[
X_{3}^{e}=\varphi\partial_{\varphi}+s\partial_{s}+t\partial_{t}+2\epsilon\partial_{\epsilon},\,\,\,X_{4}^{e}=t\partial_{t}-2g\partial_{g},
\]
\[
X_{5}^{e}=\partial_{\varphi},\,\,\,X_{6}^{e}=\partial_{s},\,\,\,X_{7}^{e}=\partial_{t}.
\]

In the group classification we use the transformations corresponding
to the generators $X_{1}^{e}$ and $X_{2}^{e}$ which are
\[
\tilde{\varphi}=e^{a}\varphi,\,\,\,\tilde{t}=e^{a}t,\,\,\,\tilde{s}=e^{4a}s,\,\,\,\tilde{H}_{b}=e^{2a}H_{b}
\]
and
\[
\tilde{H}_{b}=H_{b}+a,
\]
where $a$ is a group parameter and only changeable variables are presented.
The generators $X_{3}^{e}$ and $X_{4}^{e}$ allow reducing the constants
$g$ and $\epsilon$ to the simple case $g=1$ and $\epsilon=1$.

It should be noted that the Galilean transformation corresponding
to the generator $X_{9}^{e}=t\partial_{\varphi}$ is absent among
the equivalence transformations. This is related with the property
that the bottom does not depend on time $t$.

There are also the obvious involutions
\[
E_{1}:\,\,\,\,\tilde{t}=-t,
\]
\[
E_{2}:\,\,\,\,\tilde{\varphi}=-\varphi,\,\,\,\tilde{s}=-s.
\]

Calculations show that the kernel of admitted Lie algebras is defined
by the generators

\[
\partial_{t},\,\,\,\partial_{s}.
\]
An extension of the kernel only occurs for a linear bottom
\begin{equation}
H_{b}=qx+\beta,\label{eq:flat_topography}
\end{equation}
where without loss of generality one can assume that $\beta=0$. The
extension is defined by the generators
\[
\partial_{\varphi},\,\,\,t\partial_{\varphi},\,\,\,t\partial_{t}+4s\partial_{s}+2\varphi\partial_{\varphi}.
\]

\begin{remark} The same Lie group is admitted by the equations with
the horizontal bottom ($q=0$) studied in \cite{bk:SiriwatKaewmaneeMeleshko2016}.
 The coincidence of the admitted
Lie groups proposes to assume that the Green-Naghdi equations with
a flat bottom topography (\ref{eq:flat_topography}) are equivalent
to the Green-Naghdi equations with a horizontal bottom $q=0$. We
have checked that if such a transformation exists, then it is not
among the point transformations of the form
\[
\bar{h}=f_{1}(t,x)u+f_{2}(t,x)h+f_{0}(t,x),\,\,\,\bar{u}=g_{1}(t,x)u+g_{2}(t,x)h+g_{0}(t,x),
\]
\[
\bar{x}=\gamma_{1}(t,x),\,\,\,\bar{t}=\gamma_{2}(t,x).
\]
We have also checked that it is not among the point transformations
of the form
\[
\bar{\varphi}=f_{1}(t,s)\varphi+f_{0}(t,s),\,\,\,\bar{s}=\gamma_{1}(t,s),\,\,\,\bar{t}=\gamma_{2}(t,s).
\]
\end{remark}

\begin{remark} Equation (\ref{eq:june19.1}) with flat bottom topography
(\ref{eq:flat_topography}) have the same representations of invariant
solutions as the Green-Naghdi equations with a horizontal bottom $q=0$
given in \cite{bk:SiriwatKaewmaneeMeleshko2016}. \end{remark}

\subsection{Conservation laws }

A conservation law of equations either (\ref{eq:SGN}),
~(\ref{eq:original_bottom}) or (\ref{eq:SGN}),
~(\ref{eq:simplified_bottom}) in Lagrangian coordinates is
considered in the following local form
\[
D_{t}T^{t}+D_{s}T^{s}=0,
\]
where the densities $T^{t}$ and $T^{s}$ depend on
\[
(t,s,\varphi,\varphi_{t},\varphi_{s},\varphi_{tt},\varphi_{ts},\varphi_{ss},\varphi_{ttt},\varphi_{tts},\varphi_{tss},\varphi_{sss}).
\]
Its counterpart in Eulerian coordinates has the form
\[
D_{t}{}^{e}T^{t}+D_{x}T^{x}=0,
\]
where
\begin{equation}
^{e}T^{t}=hT^{t},\ \ T^{x}=huT^{t}+T^{s}.\label{conserved vec_in_Shall Eulerian Coor}
\end{equation}

Notice that if a generator
$X=\xi^{t}\partial_{t}+\xi^{s}\partial_{s}+\zeta\partial_{\varphi}$
is either variational or divergently variational, then there exist
such functions $B^t$ and $B^s$ that \cite{bk:Ibragimov2011}
\[
X{\cal L}+{\cal L}(D_{t} \xi^{t}+D_{s}\xi^{s}) = D_tB^t+D_sB^s.
\]
Hence, for either variational or divergently variational generator $X$, one has that
\begin{equation}
\frac{\delta}{\delta\varphi}\left(X{\cal L}+{\cal L}(D_{t}
\xi^{t}+D_{s}\xi^{s})\right)=0.
\label{eq:june25.1}
\end{equation}

\subsubsection{Conservation laws corresponding to the kernel of admitted Lie algebras}

Conservation laws are obtained by applying Noether's theorem.

The generator $\partial_{s}$ provides the conservation law with the
densities:
\[
T_{1}^{t}=\epsilon\varphi_{s}^{-3}\varphi_{tss}-3\varphi_{t}(2\varphi_{s}(1+\epsilon H_{b}^{\prime}{}^{2})+\epsilon H_{b}^{\prime\prime})+\epsilon\varphi_{s}^{-4}\varphi_{ss}(6H_{b}^{\prime}\varphi_{t}\varphi_{s}^{2}-5\varphi_{ts}),
\]
\[
T_{1}^{s}=\epsilon\varphi_{s}^{-3}\varphi_{tts}+3\varphi_{t}^{2}(1+\epsilon H_{b}^{\prime}{}^{2})-6\varphi_{s}^{-2}(\epsilon H_{b}^{\prime}\varphi_{t}\varphi_{ts}+g\varphi_{s})-6gH_{b}.
\]
In Eulerian coordinates the counterpart conservation law has the
densities
\[
^{e}T_{1}^{t}=\epsilon h^{2}v_{xx}+4\epsilon hh_{x}v_{x}-6v-3\epsilon v(H_{b}^{\prime\prime}h+2H_{b}^{\prime}{}^{2}+2H_{b}^{\prime}h_{x}),
\]
\[
T_{1}^{x}=\epsilon h^{2}(v_{tx}+v_{x}^{2})+2\epsilon h^{2}vv_{xx}+2\epsilon hvv_{x}(2h_{x}-3H_{b}^{\prime})-6g(H_{b}+h)-3v^{2}(1+H_{b}^{\prime\prime}\epsilon h+H_{b}^{\prime}{}^{2}\epsilon+2H_{b}^{\prime}\epsilon h_{x}).
\]

The generator $\partial_{t}$ gives the conservation law of energy with the
densities:
\[
T_{2}^{t}=-6\varphi_{t}^{2}(1+\epsilon H_{b}^{\prime}{}^{2})+\varphi_{s}^{-5}(2\epsilon\varphi_{t}\varphi_{tss}\varphi_{s}-8\epsilon\varphi_{t}\varphi_{ts}\varphi_{ss}-6\varphi_{s}^{4}g-3H_{b}^{\prime\prime}\epsilon\varphi_{t}^{2}\varphi_{s}^{4}+6H_{b}^{\prime}\epsilon\varphi_{t}^{2}\varphi_{s}^{2}\varphi_{ss}+12Q_{b}\varphi_{s}^{3}\varphi_{ss}),
\]
\[
T_{2}^{s}=\varphi_{s}^{-4}(2\epsilon\varphi_{t}\varphi_{tts}-2\epsilon\varphi_{tt}\varphi_{ts}-6\varphi_{t}\varphi_{s}^{2}g-3H_{b}^{\prime\prime}\epsilon\varphi_{t}^{3}\varphi_{s}^{2}-6H_{b}^{\prime}\epsilon\varphi_{t}^{2}\varphi_{ts}\varphi_{s}-12Q_{b}\varphi_{ts}\varphi_{s}^{2}-12\varphi_{t}\varphi_{s}^{3}gH_{b}).
\]
In Eulerian coordinates the densities of the corresponding conservation
law are
\[
^{e}T_{2}^{t}=\frac{1}{2}(-3H_{b}^{\prime\prime}\epsilon h^{2}v^{2}-6H_{b}^{\prime}{}^{2}\epsilon hv^{2}-6H_{b}^{\prime}\epsilon hh_{x}v^{2}-12Q_{b}h_{x}+2\epsilon h^{3}vv_{xx}+6\epsilon h^{2}h_{x}vv_{x}-6gh^{2}-6hv^{2}),
\]
\[
\begin{array}{c}
T_{2}^{x}=-3H_{b}^{\prime\prime}\epsilon h^{2}v^{3}-3H_{b}^{\prime}{}^{2}\epsilon hv^{3}-3H_{b}^{\prime}\epsilon h^{2}v^{2}v_{x}-3H_{b}^{\prime}\epsilon hh_{x}v^{3}-6Q_{b}hv_{x}-6Q_{b}h_{x}v+2\epsilon h^{3}v^{2}v_{xx}+\epsilon h^{3}vv_{tx}\\
-\epsilon h^{3}v_{t}v_{x}+3\epsilon h^{2}h_{x}v^{2}v_{x}-6gH_{b}hv-6gh^{2}v-3hv^{3}.
\end{array}
\]

\subsubsection{Flat bottom topography }

In this case one has that $H_{b}=qx$.

Using the general solution for a Lagrangian, it can be shown that
the generator $X=t\partial_{t}+4s\partial_{s}+2\varphi\partial_{\varphi}$
does not satisfy the condition (\ref{eq:june25.1}).

The generator $t\partial_{\varphi}$ corresponding to the Galilean
transformation provides the conservation law  with the densities:
\[
T_{1}^{t}=3gqt^{2}+6(1+\epsilon q^{2})(t\varphi_{t}-\varphi)-3\epsilon\varphi_{s}^{-1}q+t\varphi_{s}^{-5}(2\epsilon\varphi_{ss}(\varphi_{ts}-3q\varphi_{t}\varphi_{s}^{2})-\epsilon\varphi_{tss}\varphi_{s}),
\]
\[
T_{1}^{s}=\varphi_{s}^{-4}(t(6\epsilon\varphi_{t}\varphi_{ts}\varphi_{s}q-\epsilon\varphi_{tts}+3\varphi_{s}^{2}g)+\epsilon\varphi_{ts}-3\epsilon\varphi_{t}\varphi_{s}^{2}q).
\]
In Eulerian coordinates this conservation law gives a the center-of-mass law:
\[
^{e}T_{1}^{t}=h(6(\epsilon q ^{2}+1)(tv-x)-3\epsilon q h+6\epsilon q h_{x}tv-\epsilon h^{2}tv_{xx}-3\epsilon hh_{x}tv_{x}+3gq t^{2}),
\]
\[
\begin{array}{c}
T_{1}^{x}=h(6\epsilon q ^{2}tv^{2}-6\epsilon q ^{2}vx+6\epsilon q htvv_{x}-6\epsilon q hv+6\epsilon q h_{x}tv^{2}-2\epsilon h^{2}tvv_{xx}-\epsilon h^{2}tv_{tx}-\epsilon h^{2}tv_{x}^{2}+\epsilon h^{2}v_{x}\\
-3\epsilon hh_{x}tvv_{x}+3gq t^{2}v+3ght+6tv^{2}-6vx).
\end{array}
\]

The generator $\partial_{\varphi}$ provides the conservation law
with the densities:
\[
T_{2}^{t}=6\varphi_{t}(1+\epsilon q^{2})+6gqt+\varphi_{s}^{-5}(2\epsilon\varphi_{ss}(2\varphi_{ts}-3q\varphi_{t}\varphi_{s}^{2})-\epsilon\varphi_{tss}\varphi_{s}),
\]
\[
T_{2}^{s}=\varphi_{s}^{-4}(6\epsilon\varphi_{t}\varphi_{ts}\varphi_{s}q-\epsilon\varphi_{tts}+3\varphi_{s}^{2}g),
\]
which in Eulerian coordinates corresponds the conservation law of momentum with the densities:
\[
^{e}T_{2}^{t}=h(6\epsilon q^{2}v+6\epsilon qh_{x}v-\epsilon h^{2}v_{xx}-3\epsilon hh_{x}v_{x}+6gqt+6v),
\]
\[
T_{2}^{x}=h(6\epsilon q^{2}v^{2}+6\epsilon qhvv_{x}+6\epsilon qh_{x}v^{2}-2\epsilon h^{2}vv_{xx}-\epsilon h^{2}v_{tx}-\epsilon h^{2}v_{x}^{2}-3\epsilon hh_{x}vv_{x}+6gqtv+3gh+6v^{2}).
\]

\section{Equations (\ref{eq:SGN}), (\ref{eq:simplified_bottom}) in Lagrangian
coordinates}

\subsection{Lagrangian coordinates}

Similar analysis of equations (\ref{eq:SGN}), (\ref{eq:original_bottom})
gives that in mass Lagrangian coordinates equations (\ref{eq:SGN}),
(\ref{eq:simplified_bottom}) are equivalent to the equation
\begin{equation}\label{eq:june27.1}
  \frac{\epsilon}{3}\left(\varphi_{ttss}\varphi_{s}^{2}-4\varphi_{s}(\varphi_{ss}\varphi_{tts}
+\varphi_{ts}\varphi_{tss})+10\varphi_{ts}^{2}\varphi_{ss}\right)
-\varphi_{s}^{3}(\varphi_{tt}\varphi_{s}^{3}-g\varphi_{ss})
=g\epsilon\varphi_{s}^{6}H_{b}^{\prime}.
\end{equation}

Following the same method of finding a Lagrangian for equation (\ref{eq:june27.1}) as in the previous section, we derived the general form of the Lagrangian. A particular form of the Lagrangian providing equation (\ref{eq:june27.1}) as the Euler-Lagrange
equation ${\displaystyle \frac{\delta{\cal L}}{\delta\varphi}=0}$
is
\[
{\cal L}=\frac{1}{2}\left(\varphi_{t}^{2}-\frac{1}{2}g\varphi_{s}^{-1}\right)+\frac{1}{6}\epsilon\varphi_{ts}^{2}\varphi_{s}^{-4}-\epsilon gH_{b}.
\]

\subsection{Equivalence transformations}

\subsubsection{Equivalence group}

Calculations give that equivalence group is defined by the generators
\[
Y_{1}^{e}=t\partial_{t}+4s\partial_{s}+2\varphi\partial_{\varphi}+2H_{b}\partial_{H_{b}},\,\,\,X_{2}^{e}=\partial_{H_{b}},
\]
\[
Y_{3}^{e}=2t\partial_{t}+5s\partial_{s}+3\varphi\partial_{\varphi}+2\epsilon\partial_{\epsilon},\,\,\,X_{4}^{e}=t\partial_{t}-2g\partial_{g},
\]
\[
X_{5}^{e}=\partial_{\varphi},\,\,\,X_{6}^{e}=\partial_{s},\,\,\,X_{7}^{e}=\partial_{t}.
\]
As in the previous case there are also the obvious involutions
\[
E_{1}:\,\,\,\,\tilde{t}=-t,
\]
\[
E_{2}:\,\,\,\,\tilde{\varphi}=-\varphi,\,\,\,\tilde{s}=-s.
\]

\subsubsection{Flat bottom}

Consider the topography
\[
H_{b}=qx+\beta.
\]
Because of the equivalence transformation corresponding to the generator
$Y_{1}^{e}$, one can assume that $q^{2}=1$. Direct calculations
show that the change either
\[
\tilde{t}=t+\frac{1}{2}g\epsilon,\,\,\,\tilde{x}=qx+\frac{1}{2}g\epsilon t^{2}+k_{1}t+k_{2},
\]
\[
\tilde{u}=qu+g\epsilon t+k_{1},\,\,\,\tilde{h}=h,
\]
or
\[
\tilde{t}=k(t+\frac{k}{2}g\epsilon),\,\,\,\tilde{x}=k^{2}(qx+\frac{1}{2}g\epsilon t^{2})+k_{2},
\]
\[
\tilde{u}=k(qu+g\epsilon t),\,\,\,\tilde{h}=k^{2}h,
\]
reduces equations (\ref{eq:SGN}), (\ref{eq:simplified_bottom}) to
the horizontal flat bottom $\tilde{H}_{b}=0$. Here $k\neq0$, $k_{1}$
and $k_{2}$ are arbitrary constants. As the case with the horizontal
flat bottom is studied in \cite{bk:SiriwatKaewmaneeMeleshko2016},
it is excluded from our further consideration.

\begin{remark}
The property of the reduction of the shallow water equations with a flat bottom topography to the equations with the horizontal flat bottom is known for the one-dimensional hyperbolic shallow water equations \cite{bk:ChirkunovPikmullina[2014]}. For the two-dimensional hyperbolic shallow water equations it is proven in \cite{bk:Meleshko2020}.
\end{remark}

\subsection{Group classification}

The kernel of admitted Lie algebras is defined by the generators
\[
\partial_{t},\,\,\,\partial_{s}.
\]

Extensions of the kernel occur for
\[
H_{b}(x)=kx^{2}+\alpha x+\beta,\,\,\,(k\neq0),
\]
where without loss of generality one can assume that $\alpha=0$ and
$\beta=0$. The extensions depend on the sign of $k$:

if $k=\frac{q^{2}}{2\epsilon g}>0$, then the extension is given by
the generators
\[
\sin(qt)\partial_{\varphi},\,\,\,\,\cos(qt)\partial_{\varphi}.
\]

if $k=-\frac{q^{2}}{2\epsilon g}<0$, then the extension is given
by the generators
\[
e^{qt}\partial_{\varphi},\,\,\,\,e^{-qt}\partial_{\varphi}.
\]

\begin{remark}
Similar extensions of a kernel of admitted Lie groups occur for the one-dimensional hyperbolic shallow water equations with a parabolic bottom topography
\cite{bk:KaptsovMeleshko2019_2,bk:DorKapSW2019,bk:DorKapSW2020}. In the case of two-dimensional hyperbolic shallow equations with a constant Coriolis parameter such type of extensions also occur for a  circular paraboloid  \cite{bk:LeviNucciRogersWinternitz,bk:Meleshko2020}. The author of \cite{bk:Chesnokov2011} noted that for a circular parabolic bottom the admitted Lie algebra found in \cite{bk:LeviNucciRogersWinternitz} is isomorphic to the Lie algebra admitted by the classical shallow water equations with a horizontal bottom $H_b = const$ and zero Coriolis parameter. In  \cite{bk:Chesnokov2011,bk:Meleshko2020} it is proven that two-dimensional hyperbolic shallow equations with a constant Coriolis parameter with a circular parabolic bottom topography are locally equivalent to the classical gas dynamics equations.
\end{remark}

\subsection{Conservation laws }

\subsubsection{Kernel of admitted Lie algebras}

The kernel of admitted Lie algebras $\partial_{s}$ and $\partial_{t}$
gives the conservation laws with the densities, respectively:

\[
T_{1}^{t}=\epsilon\varphi_{s}^{-4}(\varphi_{tss}\varphi_{s}-5\varphi_{ts}\varphi_{ss})-6\varphi_{t}\varphi_{s},
\]
\[
T_{1}^{s}=\varphi_{s}^{-3}(\epsilon\varphi_{tts}-6\varphi_{s}^{2}g)+3\varphi_{t}^{2}-6\epsilon gH_{b}.
\]
and
\[
T_{2}^{t}=\epsilon\varphi_{s}^{-5}\varphi_{t}(\varphi_{tss}\varphi_{s}-4\varphi_{ts}\varphi_{ss})-3(2\epsilon gH_{b}+\varphi_{t}^{2}+\varphi_{s}^{-1}g),
\]
\[
T_{2}^{s}=\varphi_{s}^{-4}(\epsilon\varphi_{t}\varphi_{tts}-\epsilon\varphi_{tt}\varphi_{ts}-3g\varphi_{t}\varphi_{s}^{2}).
\]
Counterparts of these conservation laws in Eulerian coordinates have
the densities:
\[
^{e}T_{1}^{t}=\epsilon h^{2}v_{xx}+4\epsilon hh_{x}v_{x}-6v,
\]
\[
T_{1}^{x}=\epsilon h^{2}(v_{tx}+v_{x}^{2})+2\epsilon h^{2}vv_{xx}+4\epsilon hh_{x}vv_{x}-6gh-3v^{2}-6\epsilon gH_{b},
\]
and
\[
^{e}T_{2}^{t}=h(\epsilon h^{2}vv_{xx}+3\epsilon hh_{x}vv_{x}-3v^{2}-3gh-6\epsilon gH_{b}),
\]
\[
T_{2}^{x}=h(\epsilon h^{2}v(2vv_{xx}+v_{tx})-\epsilon h^{2}v_{t}v_{x}+3\epsilon hh_{x}v^{2}v_{x}-6ghv-3v^{3}-6\epsilon gH_{b}v).
\]

\subsubsection{Parabolic bottom topography}

Using the general solution for a Lagrangian, it can be shown that
the generators $e^{qt}\partial_{\varphi}$ and $e^{-qt}\partial_{\varphi}$
corresponding to the case $k=-q^{2}/(2\epsilon g)$ do not satisfy
the condition (\ref{eq:june25.1}). The other two generators $\sin(qt)\partial_{\varphi}$
and $\cos(qt)\partial_{\varphi}$ corresponding to the case $k=q^{2}/(2\epsilon g)$
provide the conservation laws, respectively:

\[
T_{3}^{t}=6\cos(qt)\varphi_{s}qs+\varphi_{s}^{-5}\sin(qt)(4\epsilon\varphi_{ts}\varphi_{ss}-\epsilon\varphi_{tss}\varphi_{s}+6\varphi_{t}\varphi_{s}^{5}),
\]
\[
T_{3}^{s}=6qs(\sin(qt)\varphi q-\cos(qt)\varphi_{t})+\varphi_{s}^{-4}(\epsilon\cos(qt)\varphi_{ts}q-\epsilon\sin(qt)\varphi_{tts}+3\sin(qt)\varphi_{s}^{2}g),
\]
and
\[
T_{4}^{t}=6(\cos(qt)\varphi_{t}-\sin(qt)\varphi_{s}qs)+\epsilon\cos(qt)\varphi_{s}^{-5}(4\varphi_{ts}\varphi_{ss}-\varphi_{tss}\varphi_{s}),
\]
\[
T_{4}^{s}=6qs(\cos(qt)\varphi q+\sin(qt)\varphi_{t})-\varphi_{s}^{-4}(\cos(qt)\epsilon\varphi_{tts}-3\cos(qt)\varphi_{s}^{2}g+\sin(qt)\epsilon\varphi_{ts}q).
\]

In Eulerian coordinates the counterparts of these conservation laws
contain the variable $s$, which satisfy the equations
\[
s_{t}=-hv,\ \ s_{x}=h.
\]
These conservation laws have the densities:
\[
^{e}T_{3}^{t}=6\cos(qt)sq+\sin(qt)h(6v-\epsilon h^{2}v_{xx}-3\epsilon hh_{x}v_{x}),
\]
\[
T_{3}^{x}=\cos(qt)\epsilon qh^{3}v_{x}+\sin(qt)(6sq^{2}x+6hv^{2}-2\epsilon h^{3}vv_{xx}-\epsilon h^{3}v_{tx}-\epsilon h^{3}v_{x}^{2}-3\epsilon h^{2}h_{x}vv_{x}+3gh^{2}),
\]
and
\[
^{e}T_{4}^{t}=\cos(qt)h(\epsilon h^{2}v_{xx}+3\epsilon hh_{x}v_{x}-6v)+6\sin(qt)sq,
\]
\[
T_{4}^{x}=\cos(qt)(2\epsilon h^{3}vv_{xx}+\epsilon h^{3}v_{tx}+\epsilon h^{3}v_{x}^{2}+3\epsilon h^{2}h_{x}vv_{x}-3gh^{2}-6sq^{2}x-6hv^{2})+\sin(qt)\epsilon qh^{3}v_{x}.
\]

\section{Preliminary analysis of the Green-Naghdi equations for constructing finite-difference schemes}

\subsection{Conservative form of the equations in Lagrangian variables}

Before construction of finite-difference schemes
we rewrite the Green-Naghdi equations for a horizontal bottom topography
\begin{equation}\label{GNwithGamma}
2\gamma\left(x_{s }^{2}x_{ttss}-4x_{s}x_{ss}x_{tts}-4x_{s }x_{ts}x_{tss}+10x_{ts}^{2}x_{ss}\right)
+x_{s}^{3}\left(2
x_{ss}-x_{s}^{3}x_{tt}\right)=0,
\end{equation}
where the following notations are used: $\varphi = x$ and $\gamma \propto \varepsilon$.
In a conservative form the latter equation becomes
\begin{equation}\label{GNdivForm}
    D_t(x_{t}) + D_s\left(
        \frac{1}{x_s^2}
    \right)
    - 2\gamma D_s\left(
        \frac{x_s x_{tts} - 2 x_{ts}^2}{x_s^5}
    \right) = 0,
\end{equation}
where it is assumed that $x_s \neq 0$.

The local conservation laws of momentum, energy, and center-of-mass law
can be rewritten as follows
\begin{equation} \label{locCLmom}
  \partial_x: \qquad D_t(x_{t}) + D_s\left(
        \frac{1}{x_s^2}
        - \frac{2\gamma (x_s x_{tts} - 2 x_{ts}^2)}{x_s^5}
    \right) = 0,
\end{equation}
\begin{equation} \label{locCLcm}
  t\partial_x: \qquad D_t(t x_{t} - x) + D_s\left(
        t\left(\frac{1}{x_s^2}
        - \frac{2\gamma (x_s x_{tts} - 2 x_{ts}^2)}{x_s^5}
    \right)\right) = 0,
\end{equation}
\begin{equation} \label{locCLenergy}
  \partial_t: \qquad
  D_t\left(
        \frac{x_t^2}{2} + \frac{1}{x_s}  + \gamma \frac{x_{ts}^2}{x_s^4}
  \right)
  + D_s\left(
        x_t \left(\frac{1}{x_s^2}
        - \frac{2\gamma (x_s x_{tts} - 2 x_{ts}^2)}{x_s^5}
    \right)\right) = 0.
\end{equation}

The conservation law of mass automatically follows from the symmetry of second derivatives~$x_{ts}=x_{st}$.

\subsection{The Green-Naghdi equations in hydrodynamic variables}

Here we represent the Green-Naghdi equations in hydrodynamic variables~$u(t,s)$, $\rho(t,s)$,
where $\rho$\footnote{We use $\rho$ instead of $h$, as $h$ is usually accepted in numerical analysis for  the finite-difference step.} is the depth of the fluid over the bottom and $s$, as before, is the mass Lagrangian coordinate.

It turns out to be especially convenient to use hydrodynamic
variables in finite-difference space~\cite{bk:DorKapSW2019}.
This allows transition from a three-layer finite-difference schemes to a two-layer ones.

The shallow water equations in hydrodynamic variables are obtained by the change
\begin{equation} \label{hydrotransform}
  x_s = \frac{1}{\rho},
  \qquad
  x_t = u.
\end{equation}
The Green-Naghdi equations become
\begin{equation} \label{hydro1}
\left(\frac{1}{\rho}\right)_t = u_s,
\end{equation}
\begin{equation} \label{hydro2}
u_t + 2\rho\rho_s = 2\gamma \left(\rho^4 (u_{ts} - 2 \rho u_s^2)\right)_s,
\end{equation}
where the first equation follows from the symmetry of second-order derivatives of the variable~$x$.

The variable~$p(t,s)$ which was introduced in~\cite{bk:Ovsiannikov[2003]} by the relation
\begin{equation}\label{pressure}
p = \rho^2,
\end{equation}
allows one to rewrite~(\ref{hydro2}) as follows
\begin{equation} \label{hydro2p}
u_t + p_s = 2\gamma \left(p^2 (u_{ts} - 2 \sqrt{p} u_s^2)\right)_s.
\end{equation}

The local conservation laws of momentum~(\ref{locCLmom}), energy~(\ref{locCLenergy}), and
center-of-mass law~(\ref{locCLcm})  become
\begin{equation}
  D_t(u) + D_s\left(
        \rho^2
        - 2\gamma p^2(u_{ts} - 2\sqrt{p} u_s^2)
    \right) = 0,
\end{equation}
\begin{equation}
  D_t(t u - x) + D_s\left(
        t\left[ \rho^2
        - 2\gamma p^2(u_{ts} - 2\sqrt{p} u_s^2)
    \right]\right) = 0,
\end{equation}
\begin{equation}
  D_t\left(
        \frac{u^2}{2} + \frac{p}{\rho} + \gamma (p\rho u_s)^2
  \right)
  + D_s\left(
        u \left[ \rho^2
        - 2\gamma p^2(u_{ts} - 2\sqrt{p} u_s^2)
        \right]
    \right) = 0.
\end{equation}


\section{Invariant difference schemes}
\label{sec:schemes}

\subsection{Invariant conservative scheme}

\begin{figure}[ht]
\centering
\includegraphics[scale=0.6]{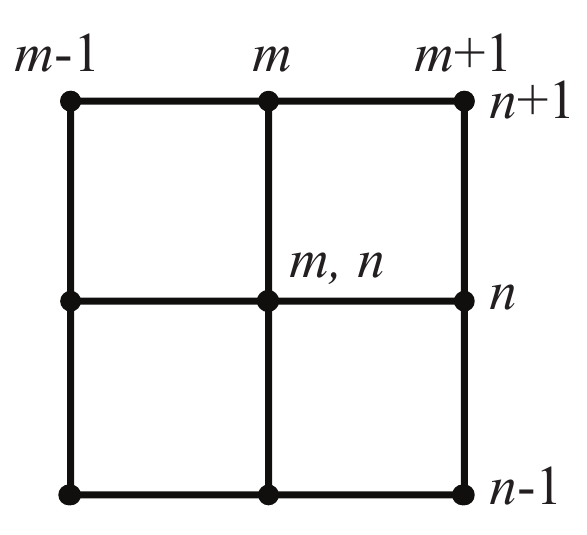}
\caption{9-point stencil}
\label{fig:template_9pt}
\end{figure}

For constructing difference scheme one needs
9-nods stencil~(see Figure~\ref{fig:template_9pt}),
which has three-layers in time
\begin{equation}\label{9ptspaceIdx}
(\mathbf{t},\mathbf{s},\mathbf{x}) =
  (
        t_n, t_{n-1}, t_{n+1};
        s_m, s_{m-1}, s_{m+1};
        x^n_m, x^n_{m+1}, x^{n+1}_{m},
        x^{n+1}_m, x^{n+1}_{m+1}, x^{n+1}_{m-1},
        x^{n-1}_m, x^{n-1}_{m+1}, x^{n-1}_{m-1}
    ),
\end{equation}
\[
    x^{n+p}_{m+q} = x(t_{n+p}, s_{m+q}),
    \qquad
    p, q \in \mathbb{Z},
\]

\indent
Here $n$- and $m$-indices correspond to the time and space layers appropriately, and
the point $(n,m)$ is fixed. For the sake of brevity we also use the following notation~\cite{bk:SamarskyPopov_book[1992]}
\begin{equation}\label{9ptspace}
(\mathbf{t},\mathbf{s},\mathbf{x}) =
  (
        t, \check{t}, \hat{t};
        s, s_-, s_+;
        x, x_+, x_-,
        \hat{x}, \hat{x}_+, \hat{x}_-,
        \check{x}, \check{x}_+, \check{x}_-,
    )
\end{equation}
Here and further the simplest orthogonal regular mesh is considered
\begin{equation}\label{regmesh}
\def\arraystretch{1.5}
\begin{array}{l}
  \tau_- = \tau_+ = \tau = t_{n+p+1} - t_{n+p} = \textrm{const},
  \\
  h_- = h_+ = h = s_{m+q+1} - s_{m+q} = \textrm{const},
  \qquad
  p, q \in \mathbb{Z},
\end{array}
\end{equation}
which is invariant with respect to the operators admitted by the
original differential equation (see also finite-difference
invariants~(\ref{GNdeltaInvs})).

Indices of finite-difference variables are shifted by the finite-difference shift operators
$\underset{\pm\tau}{S}$ and $\underset{\pm s}{S}$ that defined as follows
\[
    \def\arraystretch{1.5}
    \begin{array}{c}
    \displaystyle
    \underset{+\tau}{S}(f(t_n, s_m)) = f(t_n + \tau, s_m) = f(\hat{t}, s) = \hat{f},
    \qquad
    \underset{-\tau}{S}(f(t_n, s_m)) = f(t_n - \tau, s_m) = f(\check{t}, s) = \check{f},
    \\
    \displaystyle
    \underset{+s}{S}(f(t_n, s_m)) = f(t_n, s_m + h) = f(t, s_+) = f_+,
    \qquad
    \underset{-s}{S}(f(t_n, s_m)) = f(t_n, s_m - h) = f(t, s_-) = f_-.
    \end{array}
\]
The finite-difference total differentiation operators are defined through the shifts as
\[
    \underset{+\tau}{D} = \frac{\underset{+\tau}{S} - 1}{t_{n+1} - t_{n}},
    \quad
    \underset{-\tau}{D} = \frac{1 - \underset{-\tau}{S}}{t_n - t_{n-1}},
    \quad
    \underset{+s}{D} = \frac{\underset{+s}{S} - 1}{s_{m+1} - s_{m}},
    \quad
    \underset{-s}{D} = \frac{1 - \underset{-s}{S}}{s_m - s_{m-1}},
\]
and the following notation is used for difference derivatives
\[
    x_t = \underset{+\tau}{D}(x),
    \qquad
    x_{\check{t}} = \underset{-\tau}{D}(x),
    \qquad
    x_{t\check{t}} = \underset{+\tau}{D}\underset{-\tau}{D}(x),
    \qquad
    x_{t\check{t}t} = \underset{+\tau}{D}\underset{-\tau}{D}\underset{+\tau}{D}(x),
\]
\[
    x_s = \underset{+s}{D}(x),
    \qquad
    x_{\bar{s}} = \underset{-s}{D}(x),
    \qquad
    x_{s\bar{s}} = \underset{+s}{D}\underset{-s}{D}(x),
    \qquad
    x_{s\bar{s}s} = \underset{+s}{D}\underset{-s}{D}\underset{+s}{D}(x),
    \qquad
    \text{etc.}
\]
Notice that the operators
~$\underset{\pm\tau}{S}$, $\underset{\pm s}{S}$,
$\underset{\pm\tau}{D}$ and $\underset{\pm s}{D}$
commute in any combination on uniform orthogonal
meshes so that the following relations are valid
\[
    \check{x}_t = x_{\check{t}},
    \qquad
    \check{x}_{tt} = x_{\check{t}t} = x_{t\check{t}},
    \qquad
    x^-_s = x_{\bar{s}},
    \qquad
    x^-_{ss} = x_{\bar{s}s} = x_{s\bar{s}},
    \qquad
    \text{etc.}
\]

On the 9-nods stencil there was constructed invariant
scheme for shallow water equations in~\cite{bk:DorKapSW2019}:
\begin{equation}\label{Shallowwater}
x_{\check{t}t}
        + \dsm{D}\left[\frac{1}{\hat{x}_s \check{x}_s}\right] = 0,
    \qquad
    \hat{\tau} = \check{\tau} = \tau = \textrm{const},
    \qquad
    h_+  = h_- = h = \textrm{const},
\end{equation}
which possesses the whole number of appropriate conservation laws.


Analysis of the scheme along with consideration
of difference invariants~(\ref{GNdeltaInvs}) suggests
the following invariant extension for equation~(\ref{Shallowwater})

\begin{equation}\label{GNschemeMain}
x_{\check{t}t}
        + \dsm{D}\left[\frac{1}{\hat{x}_s \check{x}_s}\right]
  - 2 \gamma \dsm{D}\left[
        \frac{1}{\hat{x}_s^2 \check{x}_s^2} \left(
            x_{t\check{t}s} - 2 \frac{x_{ts} x_{\check{t}s}}{x_s}
        \right)
  \right]
    = 0,
\end{equation}
\[
    \tau_- = \tau_+ = \tau = \textrm{const},
    \qquad
    h_- = h_+ = h = \textrm{const}.
\]
The scheme can also be obtained with the help of the finite-difference analog
of so-called the `direct method'~(see~\cite{bk:DorKapSW2019,bk:DorKapSW2020}
and~\cite{bk:ChevDorKap2020} for discussion in detail).

Scheme~(\ref{GNschemeMain}) approximates equation~(\ref{GNdivForm}) to the order~$O(\tau^2 + h^2)$.
It possesses the following local difference conservation laws:
\begin{enumerate}
  \item
  Conservation law of mass:
  \begin{equation} \label{lagrCLmass}
    \dtm{D}(\hat{x}_s) - \dsm{D}(x_t^+) = 0.
  \end{equation}
  It automatically follows from the commutativity
  of the finite-difference differentiation on a uniform orthogonal mesh;

  \item
  Conservation law of momentum:
  \[
    \dtm{D}\left(
        x_t
    \right)
    + \dsm{D}\left[
        \frac{1}{\hat{x}_s \check{x}_s}
        - \frac{2 \gamma}{\hat{x}_s^2 \check{x}_s^2} \left(
            x_{t\check{t}s} - 2 \frac{x_{ts} x_{\check{t}s}}{x_s}
        \right)
    \right] = 0;
  \]

  \item
  {Center-of-mass law}:
  \[
    \dtm{D}\left(
        t x_t - x
    \right)
    + \dsm{D}\left[
        t \frac{1}{\hat{x}_s \check{x}_s}
        - t \frac{2 \gamma}{\hat{x}_s^2 \check{x}_s^2} \left(
            x_{t\check{t}s} - 2 \frac{x_{ts} x_{\check{t}s}}{x_s}
        \right)
    \right] = 0;
  \]

  \item
  Conservation law of energy:
  \[
    \dtm{D}\left[
        \frac{x_t^2}{2}
            + \frac{1}{2}\left(
                \frac{1}{x_s} + \frac{1}{\hat{x}_s}
            \right)
        +\gamma \left\{\dtp{D}\left(\frac{1}{x_s}\right)\right\}^2
    \right]
    + \dsm{D}\left[
        \frac{x_t^+ + \check{x}_t^+}{2}\left\{\frac{1}{\hat{x}_s \check{x}_s}
            - \frac{2 \gamma}{\hat{x}_s^2 \check{x}_s^2} \left(
                x_{t\check{t}s} - 2 \frac{x_{ts} x_{\check{t}s}}{x_s}
            \right)
        \right\}
    \right] = 0.
  \]
\end{enumerate}

\subsection{Invariant representation of scheme~(\ref{GNschemeMain})}

Scheme~(\ref{GNschemeMain}) admits the same Lie algebra as its differential counterpart, i.e.,
the generators
\begin{equation} \label{GNAlg}
X_{1}=\partial_{t},\quad
X_{2}=\partial_{s },\quad
X_{3}=\partial_{x},\quad
X_{4}=t\partial_{x},\quad
X_{5}=t\partial_{t}+4s \partial_{s }+2x\partial_{x}.
\end{equation}
In 15-dimensional space of the difference stencil~(\ref{9ptspace}), there are $15-5=10$ difference
invariants:
\begin{equation}\label{GNdeltaInvs}
\def\arraystretch{1.75}
\begin{array}{c}
    \displaystyle
    I_1 = \frac{\tau_+}{\tau_-},
    \qquad
    I_2 = \frac{h_+}{h_-},
    \qquad
    I_3 = \frac{h_-}{\tau_-^4},
    \\
    \displaystyle
    I_4 = \frac{x_+ - x}{\tau_-^2},
    \quad
    I_5 = \frac{\check{x}_+ - \check{x}}{\tau_-^2},
    \quad
    I_6 = \frac{x - x_-}{\tau_-^2},
    \\
    \displaystyle
    I_7 = \frac{\hat{x} - \hat{x}_-}{\tau_-^2},
    \quad
    I_8 = \frac{\hat{x}_+ - \hat{x}}{\tau_-^2},
    \quad
    I_9 = \frac{\check{x} - \check{x}_-}{\tau_-^2},
    \\
    \displaystyle
    I_{10} = \frac{(\hat{x} - \check{x}) t + (x - \hat{x}) \check{t} + (\check{x} - x) \hat{t}}{\tau_-^3}.
\end{array}
\end{equation}
Using the difference  invariants~(\ref{GNdeltaInvs}), scheme~(\ref{GNschemeMain}) can be represented  as follows
\begin{multline*}
{\frac {{ I_{10}}}{{ I_1}}}
+ I_3 \left(
        \frac{I_2^2}{I_5 I_8} - \frac{1}{I_7 I_9}
    \right)
+ 2\gamma\,\frac{I_3^2}{I_1} \Bigg\{
        \frac{I_2^3}{I_8 I_5^2}
        \left(
                1 - {\frac {2\,{ I_5} }{I_4}}
                + {\frac { \left( { I_4}-{ I_5} \right) {I_1}-{ I_4}+2\,{ I_5}}{{{ I_8}}}}
         \right)
         \\
        + {\frac { 2\,{ I_9} - { I_6}}
                {{ I_6}\,{ I_7}{{ I_9}}^{2}}}
         - {\frac {\left( { I_6}-{ I_9} \right) { I_1} + 2\,{ I_9} -{ I_6}}
                {{{ I_7}}^{2}{{ I_9}}^{2}}}
\Bigg\} = 0,
\end{multline*}
\[
    I_1 = 1,
    \qquad
    \qquad
    I_2 = 1.
\]

Hence, the invariant scheme and mesh are constructed. They possess the
difference analogs of all conservation laws.

\begin{remark}
All the generators~(\ref{GNAlg}) preserve the mesh uniformness and
orthogonality~(see criterion in~\cite{bk:Dorodnitsyn[2011]}). This
requirement is necessary for constructing invariant difference
schemes on uniform orthogonal meshes, and it is often satisfied for
the hydrodynamics-type schemes in Lagrangian coordinates.
\end{remark}

\subsection{Scheme~(\ref{GNschemeMain}) in hydrodynamic variables}

Here we introduce finite-difference hydrodynamic variables through the relations constructed
in~\cite{bk:DorKapSW2019} for the shallow water difference scheme, namely
\begin{equation} \label{lagrmassRel1}
x_t = u,
\qquad
x_s = \frac{1}{\sqrt{p}},
\end{equation}
\begin{equation} \label{lagrmassRel2}
  \frac{1}{\sqrt{\check{p}}} + \frac{1}{\sqrt{p}} = \frac{2}{\check{\rho}}.
\end{equation}
The latter relation is an implicit approximation of equation~(\ref{pressure}),
which allows one to preserve difference conservation laws.

\begin{figure}[ht]
\centering
\includegraphics[scale=0.8]{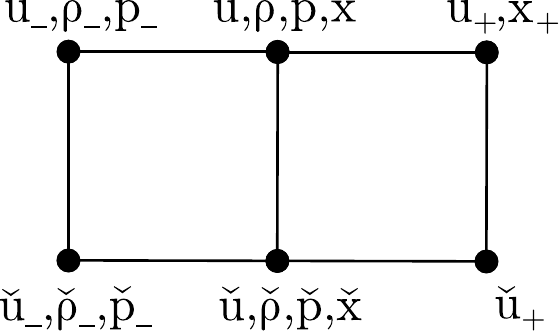}
\caption{6-point stencil}
\label{fig:template_6pt}
\end{figure}

Relations~(\ref{lagrmassRel1}) and~(\ref{lagrmassRel2}) allow one to write the scheme~(\ref{GNschemeMain})
in hydrodynamic variables on two time layers, i.e., to obtain a two-level scheme.
But at the same time the additional equations must be added to the scheme.
Hence, scheme~(\ref{GNschemeMain}) becomes
\begin{equation} \label{GNschemeMass}
    \def\arraystretch{1.75}
    \begin{array}{c}
    \displaystyle
    \dtm{D}\left( \frac{1}{\rho} \right) - \dsm{D}\left(
        \frac{u^+ + \check{u}^+}{2}
    \right) = 0,
    \\
    \displaystyle
      \dtm{D}(u) + \dsm{D}\left(
        W
    \right) = 0,
    \\
    \displaystyle
    x_t = u,
    \qquad
    x_s = \frac{1}{\sqrt{p}},
    \qquad
    \frac{1}{\sqrt{\check{p}}} + \frac{1}{\sqrt{p}}= \frac{2}{\check{\rho}},
    \\
    \displaystyle
    h_+ = h_-, \qquad
    \tau_+ = \tau_-,
    \end{array}
\end{equation}
where
\begin{equation} \label{flux_Q}
    W =
      \left[
            \frac{4}{\rho \check{\rho}}
            - \frac{2}{\sqrt{p}}\left( \frac{1}{\rho} + \frac{1}{\check{\rho}} \right)
            + \frac{1}{p}
      \right]^{-1}
      \!\!\!
      - 2 \gamma \, \frac{p \check{p} \rho^2}{(2\sqrt{p} - \rho)^2}\left(
                \check{u}_{ts} - 2 u_s\check{u}_s \sqrt{p}
            \right).
\end{equation}
The scheme is defined on 6-point stencil (see Figure~\ref{fig:template_6pt}).
One can check that equations of the system
approximate system~(\ref{hydro1}),~(\ref{hydro2p}) to the order~$O(\tau + h)$.
Notice that the first equation of system~(\ref{GNschemeMass}) follows from
the equality
\[
    \dtm{D}(x_{s} + \hat{x}_{s}) = \dsm{D}(x_{t}^+ + \check{x}_{t}^+)
\]
that is correct on the uniform orthogonal mesh.

\medskip


\bigskip

The conservation laws of  scheme~(\ref{GNschemeMass}) are the
following:
\begin{enumerate}
    \item The conservation of mass
    \begin{equation}
        {\dtm{D}}\left( \frac{1}{\rho} \right) - \dsm{D}\left(
            \frac{u^+ + \check{u}^+}{2}
        \right) = 0;
    \end{equation}

  \item The conservation law of momentum:
    \begin{equation}
        {\dtm{D}}\left( u \right) + \dsm{D}\left(W\right) = 0;
  \end{equation}

  \item The center-of-mass law:
    \begin{equation}
        {\dtm{D}}(t u - x) + \dsm{D}(t W) = 0;
    \end{equation}

  \item The conservation law of energy:
    \begin{equation} \label{scmMassEnergyCL}
        \displaystyle
        {\dtm{D}}\left(
            \frac{u^2}{2} + \frac{p}{2 \sqrt{p} - \rho}
            +  \gamma\left[\frac{p\rho u_s}{2\sqrt{p} - \rho}\right]^2
        \right)
        {} + \dsm{D}\left(
            \frac{u^+ + \check{u}^+}{2} \, W
        \right) = 0.
    \end{equation}

\end{enumerate}

\subsection{On extensions of the scheme to the case of arbitrary bottom}

For model~(\ref{eq:SGN}), (\ref{eq:simplified_bottom})
invariant scheme~(\ref{GNschemeMain}) can be extended to the case of an arbitrary bottom~$H(x)$
in the same manner as it was done in~\cite{bk:DorKapSW2019}.
Indeed,  adding
\[
\frac{{\dtm{D}}{H(x)} + {\dtp{D}}{H(x)}}{x_t + \check{x}_t}
\]
to the left-hand side of equation~(\ref{GNschemeMain}), one gets the scheme that
preserves the conservation laws of mass~(\ref{lagrCLmass}) and energy:
\[
    \dtm{D}\left[
        \frac{x_t^2}{2}
            + \frac{1}{2}\left(
                \frac{1}{x_s} + \frac{1}{\hat{x}_s}
            \right)
        +\gamma \left\{\dtp{D}\left(\frac{1}{x_s}\right)\right\}^2
        - H(x) - H(\hat{x})
    \right]
    + \dsm{D}\left[
        \frac{x_t^+ + \check{x}_t^+}{2}\left\{\frac{1}{\hat{x}_s \check{x}_s}
            - \frac{2 \gamma}{\hat{x}_s^2 \check{x}_s^2} \left(
                x_{t\check{t}s} - 2 \frac{x_{ts} x_{\check{t}s}}{x_s}
            \right)
        \right\}
    \right] = 0.
\]
The construction of an invariant scheme that conserves both energy and momentum meets considerable difficulties.
A separate example of invariant scheme that conserves momentum was proposed in~\cite{bk:DorKapSW2019}
for the shallow water equations with an arbitrary bottom.
It can be also generalized to the case of the Green-Naghdi equations.

\section{Travelling-wave type solution for the finite-difference scheme}

\subsection{Travelling-wave type solution in Lagrangian coordinates}

\begin{figure}[ht]
\centering
\includegraphics[height=120px]{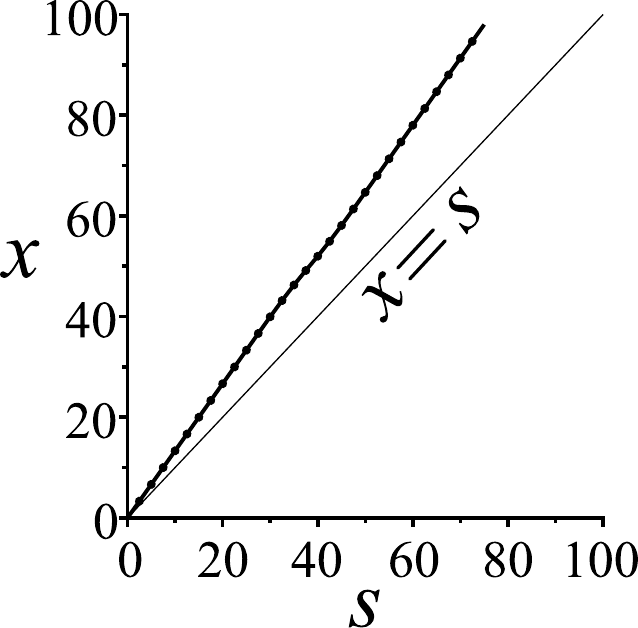}
\hspace*{2cm}
\includegraphics[height=120px]{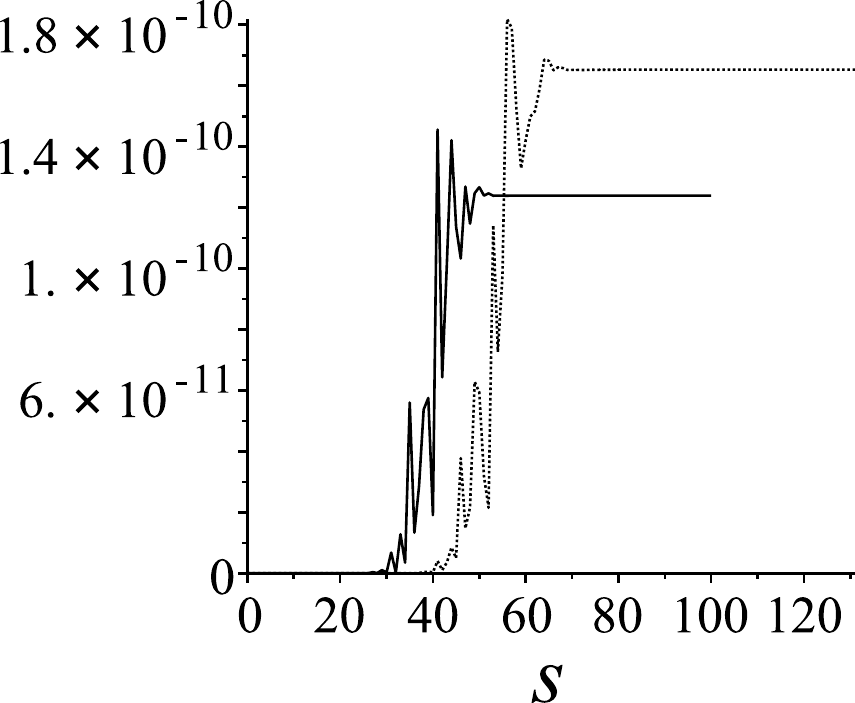}
\caption{Relation for Eulerian and mass Lagrangian coordinates for Serre's solution~($R_0 = 0.75$, $\gamma=1$).\\
Left: the solution~$x(s)$ obtained numerically.\\
Right: the numerical errors of the solution for $|s(x(s))-s|$ (solid line) and $|x(s(x))-x|$ (dotted line) confirm correctness of the relations between Eulerian coordinate~$x$ and mass Lagrangian coordinate~$s$.}
\label{fig:xs-rel}
\end{figure}

\begin{figure}[ht]
  \centering
  \includegraphics[width=0.4\linewidth]{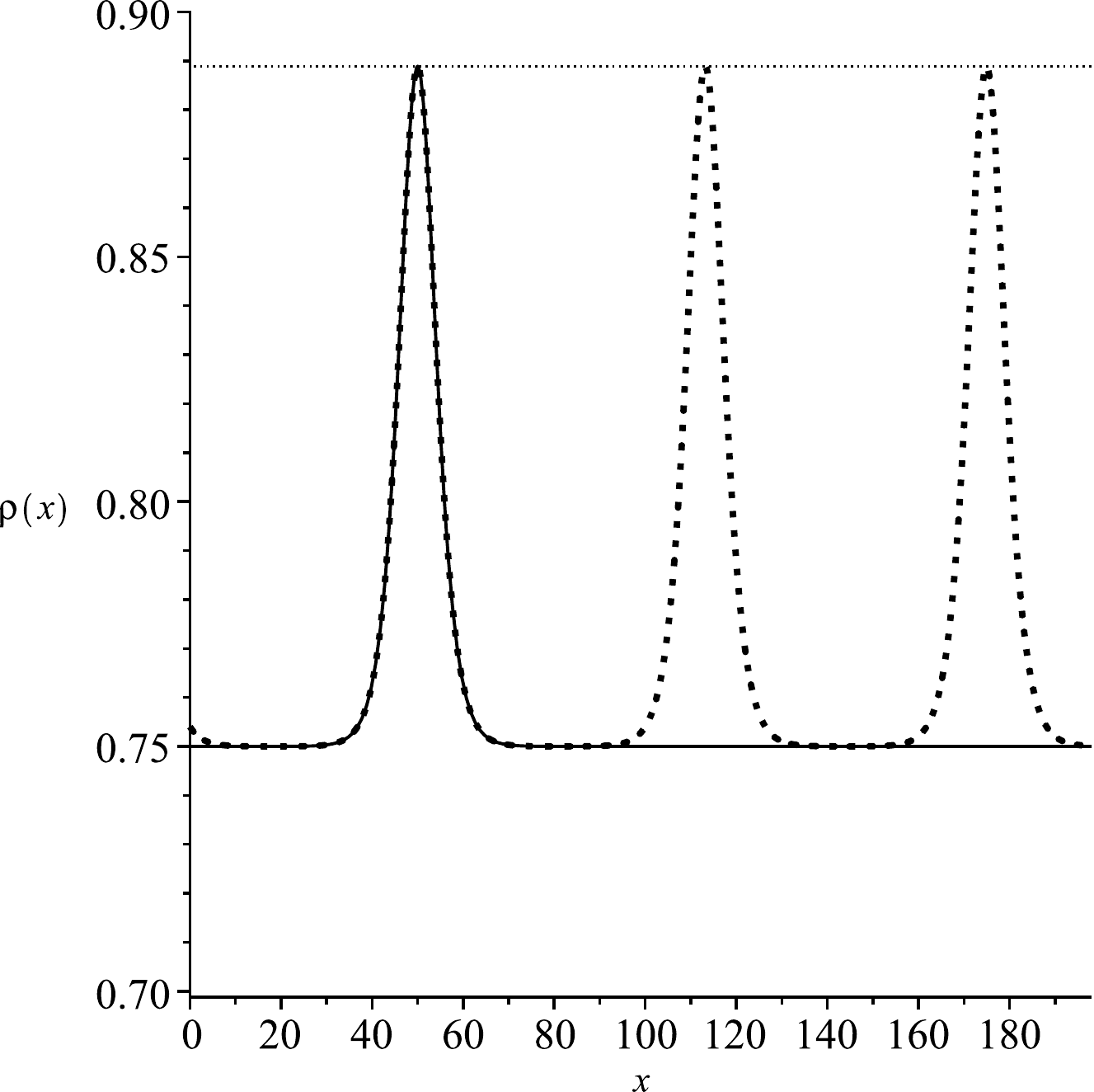}
  \hspace*{2cm}
  \includegraphics[width=0.4\linewidth]{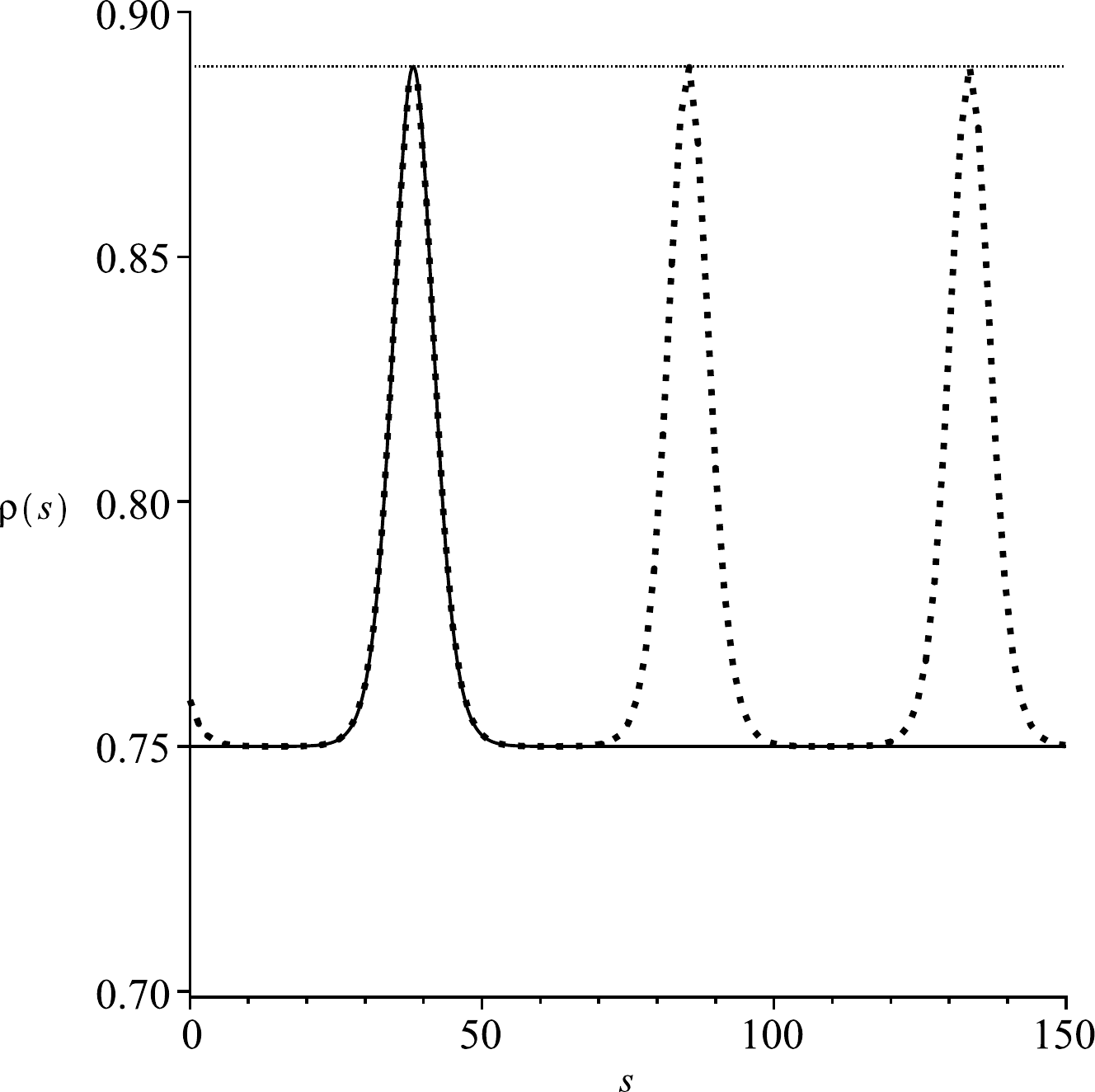}
  \caption{The Serre's solution in Eulerian~\cite{bk:SiriwatMeleshko2018} (left)
            and mass Lagrangian (center and right) coordinates
            for $R_0 = 0.75$ and $\gamma=1$.
            The data for mass Lagrangian coordinates was obtained numerically.\\
            Solid lines represents exact solutions corresponding to Serre's solution~(\ref{ExactSerre}).
            Dotted line represents numerical solutions obtained by Runge--Kutta methods.}
  \label{fig:Cuchy}
\end{figure}

To analyze some numerical properties of the constructed scheme, we
consider Serre's travelling-wave type solution~\cite{bk:Serre[1953]}
of the Green-Naghdi equations. In Eulerian coordinates, this type of
solutions is invariant with respect to the generator~$\partial_t$
and has the form\footnote{Usually the independent variable in {\bf
traveling} waves has the form~$x-ct$. Because of the Galilean
transformation the constant $c$ can be set to zero.}
\begin{equation} \label{ExactSerre}
u(x)=-\frac{k}{\rho(x)},
\qquad
\rho(x)=R_{0}+A\sech^{2}(\mu x),
\end{equation}
where
\begin{equation}\label{ExactSerreKMu}
k^2=2R_{0}^{2}(A+R_{0}),
\qquad
\mu^{2}=\frac{A}{8\gamma R_{0}^{2}(A+R_{0})}.
\end{equation}
In Lagrangian coordinates, the solution corresponds
to the set of solutions invariant with respect
to the generator~$\partial_t + \partial_s$~\cite{bk:SiriwatKaewmaneeMeleshko2016}.
It has the following representation
\[
    x(t,s) = \psi(\lambda), \qquad \lambda = s - t.
\]
The reduced equation~(\ref{GNdivForm}) is
\[
2\gamma \left( \psi^{(4)}({\psi^\prime})^2
    - 8 \psi^{\prime\prime\prime}\psi^{\prime\prime}\psi^{\prime}
    + 10 (\psi^{\prime\prime})^3
    \right)
    + \psi^{\prime\prime} (\psi^{\prime})^3
    \left(
        2 - (\psi^{\prime})^3
    \right) = 0,
\]
According to conservative form~(\ref{GNdivForm}) of the equations,
it can be rewritten as
\begin{equation} \label{GNserreRed1}
\psi^{\prime\prime}
    + D_\lambda\left((\psi^{\prime})^{-2}\right)
    - 2 \gamma D_\lambda\left(
        (\psi^{\prime})^{-4} \left(
                \psi^{\prime\prime\prime}
                - 2 \frac{(\psi^{\prime\prime})^2}{\psi^{\prime}}
            \right)
    \right) = 0.
\end{equation}
Using the change
\[
\psi^{\prime\prime} = \Theta(\zeta),
\qquad
\psi^\prime = \zeta,
\qquad
D_\lambda = \Theta D_\zeta,
\]
equation~(\ref{GNserreRed1}) becomes
\begin{equation} \label{GNserreRed2}
\left\{
    1 + D_\zeta\left(\frac{1}{\zeta^2}\right)
    - 2\gamma D_\zeta\left[
        \frac{\Theta}{\zeta^4}\left(
                \Theta^\prime - \frac{2\Theta}{\zeta}
            \right)
    \right]
\right\} \Theta = 0.
\end{equation}
If $\Theta = 0$, then one gets the trivial solution
\[
    x = \lambda.
\]
In case $\Theta \neq 0$ equation~(\ref{GNserreRed2}) has the first integral
\begin{equation}\label{GNserreRed1stIntegral}
\zeta + \frac{1}{\zeta^2}
    - 2\gamma
        \frac{\Theta}{\zeta^4}\left(
                \Theta^\prime - \frac{2\Theta}{\zeta}
            \right) = B = \textrm{const},
\end{equation}
and a particular solution corresponding to Serre's solution~(\ref{ExactSerre}) is
\begin{equation}\label{SerreTZ}
 \Theta^2(\zeta) = \frac{\zeta^3}{2\gamma}\left(\zeta - \frac{1}{R_0} \right)^2\left(\zeta - \frac{1}{R_0 + A}\right).
\end{equation}
In Eulerian coordinates it has the form
\begin{equation} \label{serreReductEuler}
    {\tilde{\rho}}_x^2 = \frac{{\tilde{\rho}}^3}{2\gamma}\left(
                \frac{1}{{\tilde{\rho}}} - \frac{1}{R_0}
            \right)^2
            \left(
                \frac{1}{{\tilde{\rho}}} - \frac{1}{R_0 + A}
            \right),
            \qquad
            {\tilde{\rho}} = {\tilde{\rho}}(t,x),
\end{equation}
and in mass Lagrangian coordinates it is
\begin{equation} \label{serreReductLagr}
    \rho_s^2 = \frac{\rho}{2\gamma}\left(
                \frac{1}{\rho} - \frac{1}{R_0}
            \right)^2
            \left(
                \frac{1}{\rho} - \frac{1}{R_0 + A}
            \right),
            \qquad
            \rho = \rho(t,s).
\end{equation}
Notice that Serre's solution satisfy the higher-order equation~(\ref{GNserreRed1})
only if~$R_0$ and~$A$ are related as follows
\[
    A = \frac{1 - 2R_0^3}{2R_0^2}.
\]
In order to perform computations in mass Lagrangian coordinates, one must find the initial distribution
which correspond to solution~(\ref{ExactSerre}).
Hence, one must solve the Cauchy problem~(\ref{lagr_mass_rel}).
For Serre's solution it has the form
\begin{equation}
\frac{1}{(x(0,s))_s} = R_0 + A \sech^2(\mu x(0, s)),
\qquad
x(0, s_0) = x_0,
\end{equation}
where
\[
    s_0 = \int_{0}^{x_0} (R_0 + A \sech^2(\mu z))dz
\]
is mass of the fluid at point $x=x_0$.
Then, the initial conditions must satisfy the following relations
\begin{equation}
  x(0, s) = \frac{z_0(s)}{\mu},
  \quad
  \text{where}
  \quad
  e^{2z_0}(\mu (s_0 - s) + A + z_0 R_0)
    + \mu (s_0 - s) - A + z_0 R_0 = 0.
\end{equation}
The solution for $R_0 = 0.75$, $\gamma=1$ is given in Figure~\ref{fig:xs-rel} (left). The center of the soliton solution is shifted to the point $x_c = 50$.
The right side of Figure~\ref{fig:xs-rel} demonstrates the numerical error of the transformations.
The numerical solutions in Eulerian~(\ref{serreReductEuler}) and mass Lagrangian~(\ref{serreReductLagr})
coordinates for~$x \in [0,180]$ are given in Figure~\ref{fig:Cuchy}.
The original analytical solution is not periodic and it is stable for small linear perturbations~\cite{bk:YiLi2001}.
However, numerical calculations implement a periodic like solution~(it was mentioned in~\cite{bk:SiriwatMeleshko2018}).
Notice that this does not contradict the linear analysis since numerical calculations introduce finite perturbations. The quasiperiodicity of the numerical solution is confirmed by calculations
with very small steps. Variation of steps from large to small does not affect the final solution sufficiently.

\subsection{Reduction of the scheme}

Reduction of the invariant scheme  on the subgroup~$\partial_t + \alpha\partial_{s}$
is the same as  the reduction for the shallow water
equations scheme~\cite{bk:DorKapSW2019}.
The reduced scheme is
\begin{equation} \label{SchemeSerreRed}
    \alpha^2 \check{\psi}_{\lambda\lambda}
    + \underset{-\lambda}D\left(
        \frac{1}{\hat{\psi}_\lambda \check{\psi}_\lambda}
    \right)
    - 2 \gamma \alpha^2 \underset{-\lambda}D\left[
        \frac{1}{(\hat{\psi}_\lambda \check{\psi}_\lambda)^2}
        \left(
            \check{\psi}_{\lambda\lambda\lambda}
            - 2 \frac{\check{\psi}_{\lambda\lambda}\psi_{\lambda\lambda}}{\psi_\lambda}
        \right)
    \right] = 0,
\end{equation}
\[
    \Delta\lambda = \hat{\lambda} - \lambda = \lambda - \check{\lambda}
    = \text{const},
\]
where $\psi = \psi(\lambda) = \psi(s - \alpha t)$.

\begin{figure}[ht]
\centering
\includegraphics[width=0.3\linewidth]{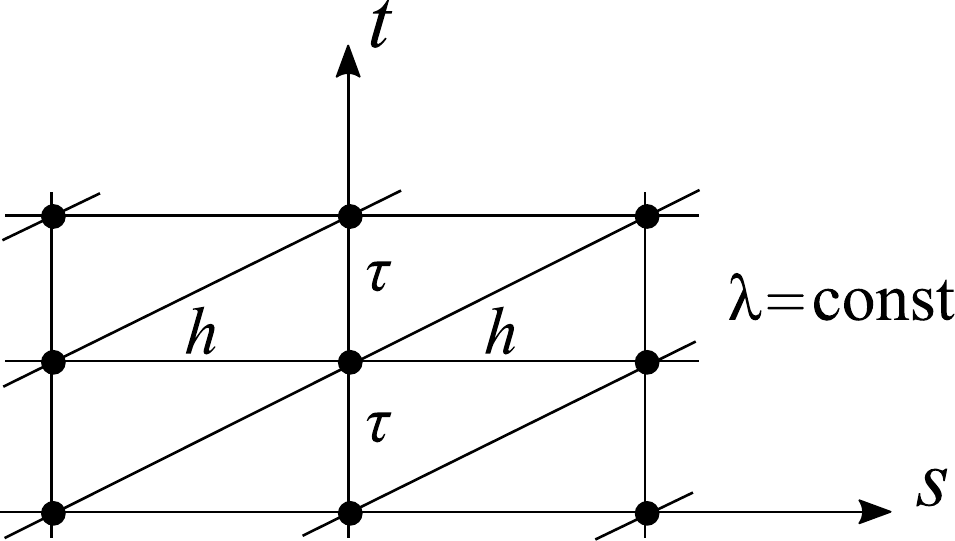}
\caption{Relation between meshes in variables $(s, t)$ and $\lambda$.}
\label{fig:meshmatch}
\end{figure}

Equation~(\ref{SchemeSerreRed}) is a finite-difference analog of equation~(\ref{GNserreRed1}).
To guarantee that new mesh spacing $\Delta\lambda$ matches the original
mesh nodes on the plane~\cite{bk:Dorodnitsyn[2011]}~(see Figure~\ref{fig:meshmatch}), one should assume that
\[
\Delta\lambda = h = \alpha\tau
\]

Below $h$ and~$\tau$ are considered equal, which corresponds to~$\alpha=1$.

\begin{remark}
It is worth mentioning that the difference shift operators in mass Lagrangian coordinates
are related to the shift operators in $\lambda$--coordinates as follows
\[
    \underset{\pm\tau}{S} = \underset{\mp\lambda}{S},
    \qquad
    \underset{\pm h}{S} = \underset{\pm \lambda}{S}.
\]
\end{remark}

\begin{figure}[ht]
    \centering
  \includegraphics[width=0.45\linewidth]{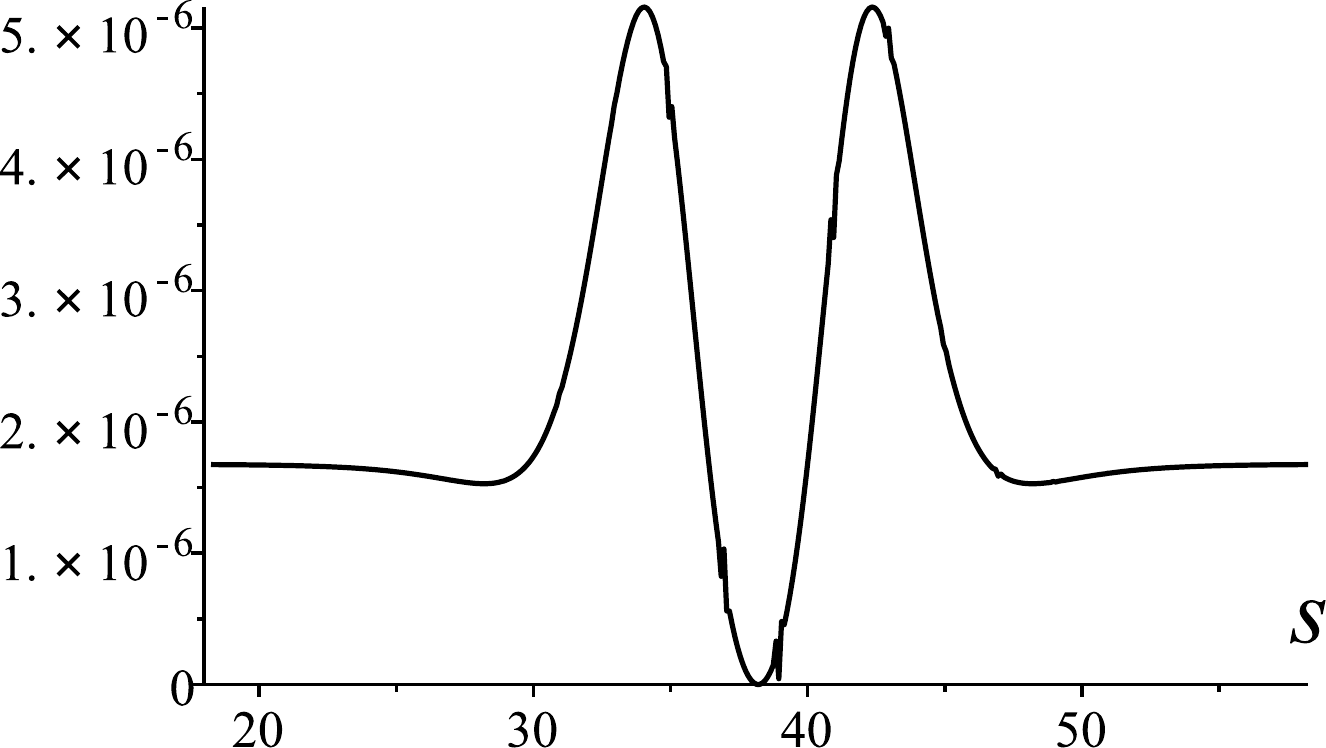}
  \caption{Preservation of the first integral~(\ref{SerreDeltaInt}) on the Serre's solution.}
  \label{fig:serre1st}
\end{figure}

\medskip
Changing the variables
\begin{equation}
    \psi_{\lambda\lambda} = \Theta(\zeta),
    \quad
    \check{\psi}_{\lambda\lambda} = \Theta(\check{\zeta}) = \check{\Theta},
    \quad
    \zeta = \psi_\lambda,
    \quad
    \underset{-\lambda}D = \check{\Theta} \underset{-\zeta}D,
\end{equation}
one gets a difference analog of ODE~(\ref{GNserreRed2}):
\begin{equation}
    \left\{
        1
        + \underset{-\zeta}D\left(\frac{1}{\hat{\zeta}\check{\zeta}}\right)
        - 2\gamma\underset{-\zeta}D\left[
            \frac{\check{\Theta}}{(\hat{\zeta}\check{\zeta})^2}\left(
                \check{\Theta}_\zeta
                - 2 \frac{\Theta}{\zeta}
            \right)
        \right]
    \right\} \check{\Theta} = 0,
\end{equation}
\[
    \Delta\zeta = \hat{\zeta} - \zeta = \zeta - \check{\zeta} = \text{const}.
\]
If $\check{\Theta} = 0$, then the appropriate solution is~$x = \lambda$
as it is in the differential case considered above.
If $\check{\Theta} \neq 0$, then one can derive the following difference first integral
\begin{equation} \label{SerreDeltaInt}
    \hat{\zeta}
    + \frac{1}{\hat{\zeta}\check{\zeta}}
    - \frac{2\gamma\check{\Theta}}{(\hat{\zeta}\check{\zeta})^2}\left(
            \check{\Theta}_\zeta
            - 2 \frac{\Theta}{\zeta}
        \right) = B_1 = \text{const},
\end{equation}
which is the finite-difference analogue of integral~(\ref{GNserreRed1stIntegral}).
The control of the values of the first integral~(\ref{SerreDeltaInt}) for  Serre's solution
near the center of the soliton for the chosen parameters is given in Figure~\ref{fig:serre1st}.
It shows high accuracy of the first integral conservation.

\begin{figure}[ht]
    \centering
  \includegraphics[width=0.45\linewidth]{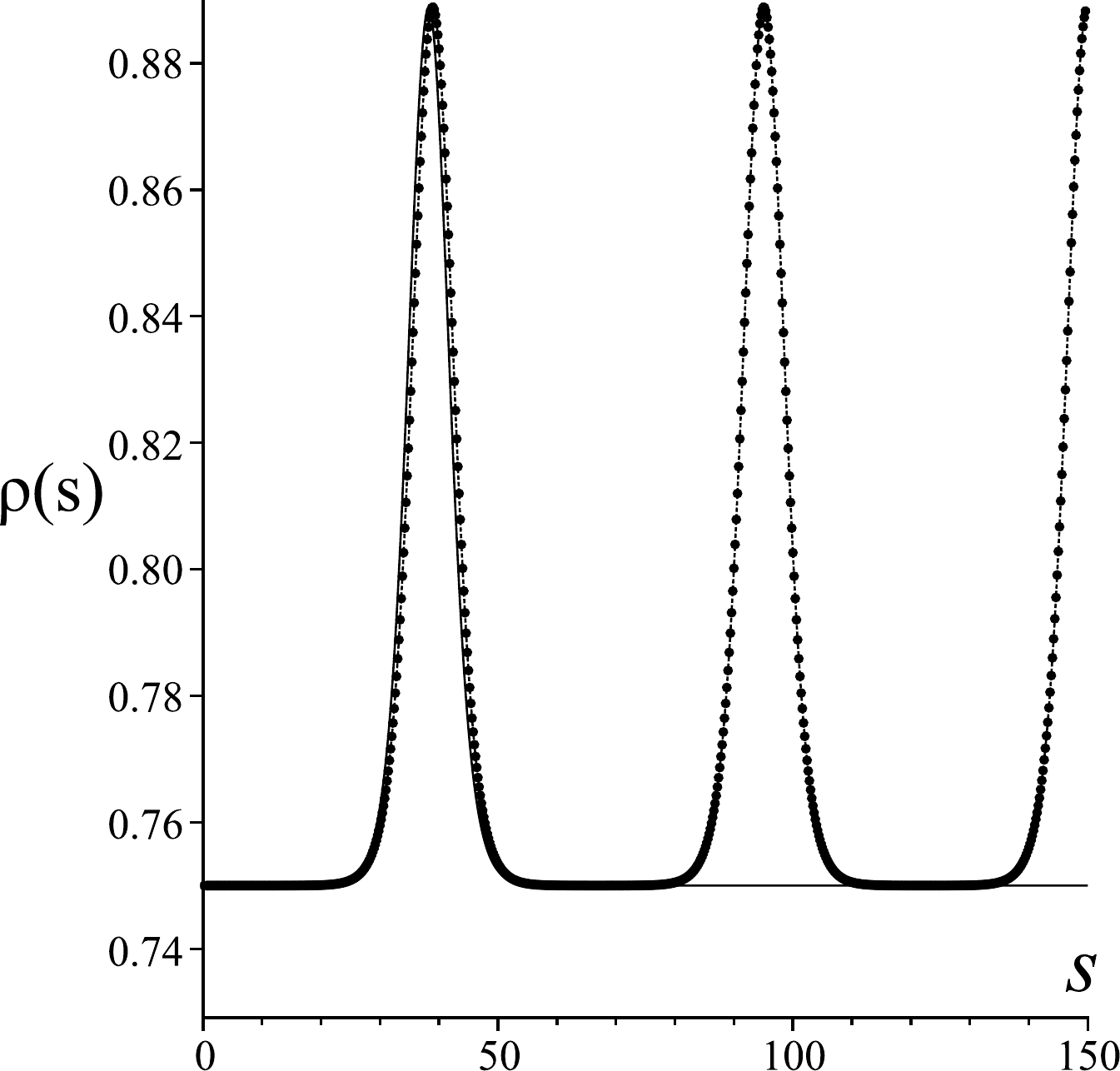}
  \caption{Numerical solution of the reduced scheme~(the initial data is given form the solution obtained by the Runge-Kutta type method).
  Solid line represents exact solution corresponding to Serre's solution~(\ref{ExactSerre}).
  Dotted line represents numerical solution of the reduced scheme.}
  \label{fig:serre_stst_sol_scheme}
\end{figure}

The solution of the reduced scheme for the initial parameters,
obtained by using Runge-Kutta method for equation~(\ref{serreReductLagr}),
is presented in Figure~\ref{fig:serre_stst_sol_scheme}.
Notice that, as in the previous section, the numerical solution is periodic like.

\subsection{Evolution of the Serre's solution for a perturbed scheme}

Consider the second equation of scheme~(\ref{GNschemeMass})
\begin{equation}\label{GNschemeMassEq2}
\dtm{D}(u) + \dsm{D}\left(
            \left[
            \frac{4}{\rho \check{\rho}}
            - \frac{2}{\sqrt{p}}\left( \frac{1}{\rho} + \frac{1}{\check{\rho}} \right)
            + \frac{1}{p}
            \right]^{-1}
      \right)
      = 2\gamma \dsm{D}\left( Q \left[
                \check{u}_{ts} - 2 u_s\check{u}_s \sqrt{p}
            \right] \right),
\end{equation}
where $Q = (p \check{p} \rho^2)/(2\sqrt{p} - \rho)^2$.

The term~$\dsm{D}(Q \check{u}_{ts}) \sim \check{u}_{ts\bar{s}}$ is proportional to~$(\tau h^2)^{-1}$
which takes on  large values. One can avoid it by putting~$\gamma \sim \tau h^2$.
But in this case the equations become the shallow water equations which are out of our interest here.
Because of the large coefficients even for very small gradients of~$u$
the scheme cannot be numerically solved by standard methods.
There are at least two known approaches to correct this situation. The
first approach~\cite{bk:DegFav68,bk:Kalitkin68} is based on the splitting the scheme into a system of difference equations
and
calculating the flow~$Q \check{u}_{ts}$ separately. It is used, for example, in magnetohydrodynamics,
where the coefficient of thermal conductivity often tends to infinity.
Another approach~\cite{bk:BonnetonLannes2011} goes beyond finite difference methods.
The hybrid scheme is constructed by splitting the Green-Naghdi equation
into two parts. The finite-difference part approximates the shallow water equation,
and the finite-volume part of the scheme approximates the $\gamma$--terms of the equation.

In order to avoid these difficulties
we consider the following perturbed version of equation~(\ref{GNschemeMassEq2})
\begin{equation} \label{GNschemeMassMod}
\dtm{D}(u) + \dsm{D}\left(
        \left[
        \frac{4}{\rho \check{\rho}}
        - \frac{2}{\sqrt{p}}\left( \frac{1}{\rho} + \frac{1}{\check{\rho}} \right)
        + \frac{1}{p}
        \right]^{-1}
  \right)
  = 2\gamma \dsm{D}\left( Q\left[
            \alpha \check{u}_{ts} - 2 \beta u_s\check{u}_s \sqrt{p}
        \right] \right),
\end{equation}
where $0 < \alpha \leqslant 1$ and $0 < \beta \leqslant 1$ are some constant values.
Tending the coefficient~$\alpha$ to zero, one can neglect
effects of the term proportional to $\check{u}_{ts\bar{s}}$.
This particular problem is also of independent interest.
To clarify the physical meaning of this case, we refer to the paper~\cite{bk:Mitsotakis2017},
where it was considered a similar to~(\ref{GNschemeMassMod}) perturbed form
of the Green-Naghdi equations in Eulerian coordinates.
The perturbation coefficient~$\alpha$ is responsible for the stability of the soliton solution.
If its value tends to zero, the solution show dispersive qualities in time. In contrast to~$\alpha$,
the value of the perturbation coefficient~$\beta$ does not essentially affect the profile of the free surface.

\begin{figure}[ht]
\centering
\includegraphics[width=0.65\linewidth]{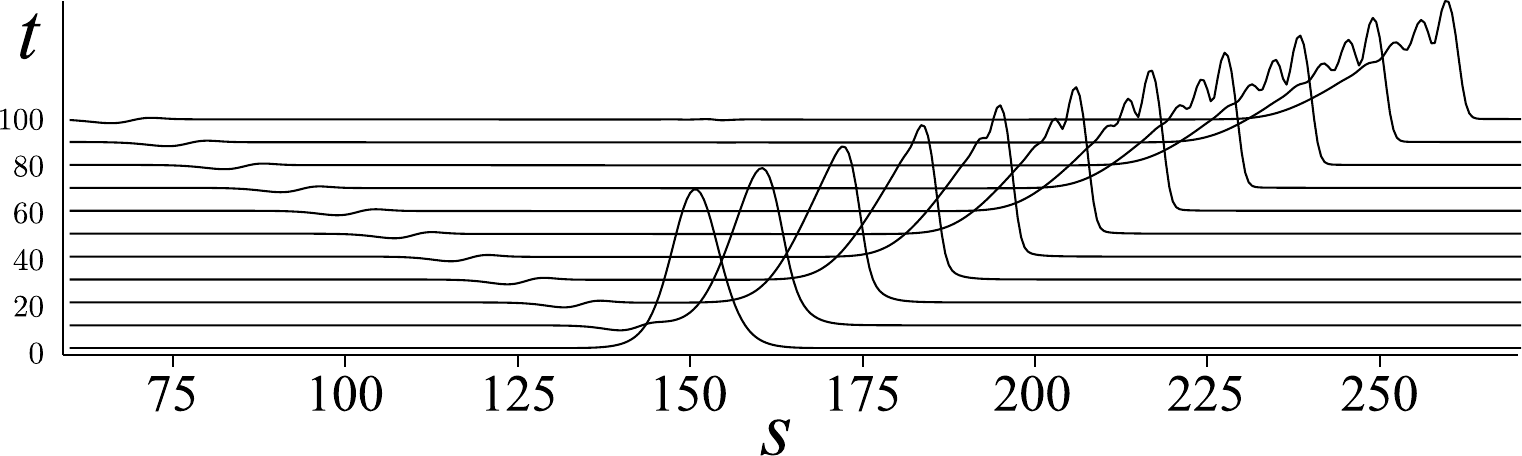}
\caption{Evolution of the Serre's solution for the perturbed scheme~(\ref{GNschemeMassMod}).}
\label{fig:serre-evol}
\end{figure}

In particular, the evolution of the Serre's solution for~$\alpha=0.001$, $\beta=1$, $R_0=0.75$ and $\gamma=1$ is presented in Figure~\ref{fig:serre-evol}.
For the represented solution we explore a viscous version of the scheme, where the following change was used
\[
 \rho_m^n \mapsto \rho_m^n - \nu\tau\!\dsm{D}(\rho_m^n),
 \qquad
 u_m^n \mapsto u_m^n - \nu\tau\!\dsm{D}(\rho_m^n u_m^n),
 \qquad
 m > 2.
\]
Here~$\nu$ is a small viscosity coefficient.
The dispersive wave forms are very similar to the results presented in~\cite{bk:Mitsotakis2017} for small values of~$\alpha$.

%

\section{Conclusion}

Group analysis of the one-dimensional Green-Naghdi equations
describing the behavior of fluid flow over uneven bottom
topography is given in the present paper. The Green-Naghdi equations
are considered in two forms: in the classical form (\ref{eq:SGN}),
(\ref{eq:original_bottom}) and in the mild slope approximation form
(\ref{eq:SGN}), (\ref{eq:simplified_bottom}) which is of the same
order as the original Green-Naghdi equations. Analysis of the
studied equations is performed in Lagrangian coordinates. Working in
the Lagrangian coordinates allowed us to find Lagrangians turning
the analyzed equations into the Euler-Lagrange equations. It is
shown that equations (\ref{eq:SGN}), (\ref{eq:simplified_bottom})
with a flat bottom topography $H_{b}=qx+\beta$ are locally
equivalent to the Green-Naghdi equations with a horizontal bottom
topography $H_{b}=\textrm{const}$. Complete group classification of
both cases of the Green-Naghdi equations with respect to the
function $H_{b}$ describing the uneven bottom topography is
presented. Applying the Noether theorem, the developed
Lagrangians and performed group classification, conservation laws
of the one-dimensional Green-Naghdi equations with uneven bottom
topography are obtained.

An invariant conservative finite-difference scheme is constructed
for the Green-Naghdi equations for the case of a flat bottom
topography. The scheme possesses the conservation laws of mass,
momentum, energy and the center-of-mass law. This scheme is also
represented in hydrodynamic variables. The representation in
hydrodynamic variables simplifies its numerical implementation. The
reduction of the invariant scheme on a subgroup is carried out
similarly to the reduction of the corresponding differential
equations. As a result of the reduction an ordinary
finite-difference equation is obtained. This equation possesses a
first integral, which is well preserved on  Serre's exact solution.
Using the example of Serre's solution  further numerical analysis of
the scheme is performed. The stationary solution, obtained by the
proposed scheme, has the same qualitative properties as the
solutions calculated by the Runge-Kutta methods. The time evolution
of Serre's solution is considered by the example of a specific
perturbed version of the scheme which allows one to avoid working
with large velocity gradients terms. The latter invariant scheme was
generalized for an arbitrary bottom topography. This scheme also
possesses the conservation laws of mass and energy.

\section*{Acknowledgements}
The research  was supported by Russian Science Foundation Grant No
18-11-00238 "Hydrodynamics-type equations: symmetries, conservation
laws, invariant difference schemes".
E.I.K. acknowledges Suranaree University of Technology for
Full-time Master Researcher Fellowship.


\end{document}